\documentclass[a4paper,11pt]{article}

\pdfoutput=1 

\usepackage{jcappub} 
\usepackage[backend=bibtex, maxnames=4, minnames=1]{biblatex}
\usepackage[T1]{fontenc} 
\usepackage{pgfplots}
\usepackage{amsmath}
\usepackage{multirow}
\usepackage{gensymb}

\DeclareUnicodeCharacter{2009}{\,} 

\title{\boldmath Investigating the effect of hadronic models on IACT images }

\author[a]{Benedetta Bruno,}
\author[a]{Rodrigo Guedes Lang,}
\author[b]{Luan Bonneau Arbeletche,}
\author[b]{Vitor de Souza,}
\author[a]{Stefan Funk}

\affiliation[a]{Friedrich-Alexander-Universit\"at Erlangen-N\"urnberg, Erlangen Centre for Astroparticle Physics, Nikolaus-Fiebiger-Str. 2, 91058 Erlangen, Germany}
\affiliation[b]{Instituto de F\'{i}sica de S\~ao Carlos, Universidade de S\~ao Paulo, Av. Trabalhador S\~ao Carlense 400, S\~ao Carlos-SP, 13566-590, Brazil}

\emailAdd{benedetta.bruno@fau.de}
\emailAdd{rodrigo.lang@fau.de}

\abstract{The predictions of hadronic interaction models for cosmic-ray induced air showers contain inherent uncertainties due to limitations of available accelerator data. This leads to differences in shower simulations using each of those models. Many studies have been carried out to track those differences by investigating the shower development or the particle content. In this work, we propose a new approach to search for discrepancies and similarities between the models, via the IACT images resulting from the observations of hadronic air showers. We use simulations of H.E.S.S. as a show-case scenario and, by investigating variables of the camera images, we find potential indicators to highlight differences between models. Number of pixels, Hillas image size, and density showed the largest difference between the models. We then further explore the (in)compatibility of the models by combining all the variables and using Boosted Decision Trees. For protons, a significant difference in the classifier output is found for \texttt{EPOS-LHC} when compared to both \texttt{QGSJET-II04} and \texttt{Sybill 2.3d}. For helium and nitrogen, \texttt{QGSJET-II04} is shown to be the outlier case. No significant differences are found for silicon and iron. The distribution of (in)compatibility between the models in the phase space of reconstructed shower parameters shows that a targeted search can be fruitful, with showers with reconstructed energies of a few TeV and reconstructed core closer to the large telescope presenting the largest power of separation. An investigation of the distribution of first interaction parameters has shown that \texttt{EPOS-LHC} and \texttt{QGSJET-II04} result in significantly different distributions of multiplicity and height of first interaction for protons and elasticity and fraction of energy carried by neutral pions for helium and nitrogen.}

\begin{document}
\maketitle
\flushbottom

\section{Introduction}
\label{sec:introduction}

Imaging Atmospheric Cherenkov Telescopes (IACTs) detect cosmic rays and gamma rays through particle showers initiated by the interaction of these particles with atoms in the atmosphere. Sensitive to the Cherenkov light emitted by relativistic charged particles in the cascade, IACTs enable for the indirect characterization of the primary particle using the properties of images registered on fast cameras. A combination of image-treatment algorithms and posterior multi-parametric analyses result in estimates of the primary particle's energy, direction, and type. Despite being mainly designed for optimal sensitivity to primary gamma rays with energies $>100$~GeV, a dominant background of cosmic rays produces the largest number of detected images. This motivates the use of IACT observatories for purposes other than measuring gamma rays, such as the study of cosmic-ray showers.

Analyses targeting the measurement of gamma-ray fluxes imply a crucial step of decontaminating event samples from showers initiated by cosmic rays (the gamma/hadron separation step). While gamma-ray-induced showers generate elliptical images on IACTs, the hadronic interactions driving the evolution of cosmic-ray showers imbue on images a morphology that is typically erratic (see e.g.~\cite{Vlk2009}). Therefore, removal of almost 100\% of the background through simple image-parametrization schemes (Hillas parameters~\cite{1985HillasParameters}) is achievable~\cite{1997FeganGammaHadron}. More complex algorithms (such as machine-learning~\cite{2019JacquemontDL,2021MienerDL, Glombitza_2023} and template techniques~\cite{1998CatTemplate,2014ParsonsTemplate}) and improvement of detector capabilities can optimize background removal. However, a natural limit exists~\cite{2007SobczyskaNaturalLimit} because the stochastic nature of hadronic interactions can result in the transference of a large fraction of the primary energy into the electromagnetic component, thus mimicking gamma-ray events~\cite{2007MaierGammaLike,2018SitarekGammaLike}. These gamma-like images are retained as background and their separation from the gamma-ray signal can only be performed statistically. Some methods include measuring off-signal samples (in the case of point-like sources)~\cite{2007BergeBackgroundModelling}, comparing with simulations and/or modeling of the cosmic-ray background (in the case of extended sources~\cite{2014AckermannFermiBubbles,2017AbeysekaraExtended,Mohrmann:2019hfq} and, beyond, in measuring the diffuse electron flux~\cite{2008HESSElectron,2018VeritasElectron}). Moreover, the need for simulations of cosmic ray events is indispensable when designing and estimating sensitivities of future observatories~\cite{2019CtaOptimization}.

Importantly, in cases where the simulation of cosmic-ray showers is required, it has been found that modeling hadronic interactions is a major source of systematic uncertainties, mostly due to the difference in the predicted fractions of gamma-like events. In an analysis targeting the future Cherenkov Telescope Array Observatory~\cite{2021OhishiEolHadronic} (CTAO), the estimated sensitivity to point-like gamma-ray sources has been shown to differ up to 30\% when computed with different hadronic interaction models (most notably, in the range 1~TeV~$< E <$~30~TeV). Models favoring the production of hard neutral pions result in a larger fraction of gamma-like events, and that fraction is known to be model-dependent, even in the IACT energy range~\cite{2007MaierGammaLike}. Moreover, the prediction of shower properties, such as the electromagnetic energy deposit at ground level and Cherenkov-light density at ground level, was compared between different hadronic interaction models in Refs.~\cite{2019ParsonsHadronic,2011ParsonsHadronic} for proton-induced showers, where discrepancies between models were corroborated and shown to be present even without gamma-like event selection. In this direction, it is clear that a better understanding of hadronic interactions can potentially improve gamma-ray detection using IACTs. In particular, because gamma-like events are in the tails of the parameter distribution of protons, this might generate different systematic uncertainties for the general cosmic-ray population.

The existence of different approaches for describing hadronic interactions stems from the incompleteness of our current picture of strong interactions. Because the bulk of possible processes can not be computed in terms of first principles in perturbative Quantum Chromodynamics (QCD), phenomenological models have been elaborated to complement QCD and supply complete models for hadronic and nuclear processes. These models have free parameters that are constrained by collider data~\cite{2011dEnterriaConstraintsLHC}, and are extrapolated to unmeasured kinematic regions during air-shower simulations. Exhaustive comparisons exist~\cite{2017EolDiffractive,2018CalcagniModelsCompare} showing that characteristics of particular models lead to uncertainties and discrepancies in the spectra of final-state particles, which are reflected on the simulation shower observables.

In the end, studies for astroparticle physics relying on shower simulations have to deal with the task of choosing one among many hadronic interaction models available as software packages. The most commonly applied are \texttt{EPOS-LHC}~\cite{epos}, \texttt{Sibyll~2.3d}~\cite{sibyll}, and \texttt{QGSJetII-04}~\cite{qgs}, all of which are constrained by data from the LHC~\cite{2011dEnterriaConstraintsLHC}. While for astrophysical studies the modeling of hadronic interactions enters as a source of uncertainty, it has been argued~\cite{2019MitchellMuons,2021OhishiEolHadronic} that since IACTs are sensitive to hadronic interactions, these detectors stand as a plausible scenario for probing the hadronic interactions themselves.


Therefore, in the direction of using IACT observatories to probe hadronic interactions, we propose in this study to explore the parameter space of detection (in terms of impact distance, primary energy, and image parameters) to find regions where discrepancies between the models are most relevant and can be used to constrain those models. For that, a simulation-based analysis will be performed in the context of the High Energy Stereoscopic System (H.E.S.S) experiment. H.E.S.S. is one of the currently operational IACT arrays and is located in the Khomas Highlands of Namibia at 1800~m above sea level. It consists of five telescopes, four small telescopes with 108~m$^2$ mirror area and 15~m focal length, named CT1-4, arranged in a square with 120~m side length, and a single 614~m$^2$ mirror area telescope (CT5) of 36 m focal length placed in its center~\cite{2020NatureCrab}. In this work, we use H.E.S.S. as a show-case scenario. Nevertheless, the obtained results are more general and the technique here developed can be applied to any IACT experiment.

This paper is organized as follows. In Section~\ref{sec:simulations} the techniques and details of our simulations are presented. In Section~\ref{sec:lowvariables} the difference in Hillas parameters for different models are investigated individually. In Section~\ref{sec:highvariables}, these variables are combined in a machine-learning approach leading to a single high-level variable that improves the power of separation. In Section~\ref{sec:shower} the correlation between distributions of parameters of first shower interaction and Hillas parameters are presented for each model. Finally, in Section~\ref{sec:conclusions}, the work conclusions and outlooks are discussed.

\section{Simulations}
\label{sec:simulations}
In order to investigate the effect of hadronic interaction models on IACT images, we simulated a large number of cosmic ray (CR) extensive air showers for different models using the software package CORSIKA (Cosmic Ray Simulation for Kascade, version 7.7402)~\cite{heck_1998}. Initially developed for the KASCADE experiment~\cite{ANTONI2003490}, this software is publicly available, open-source, and a standard tool for the broader astroparticle physics community. The showers were simulated using the \texttt{QGSJET-II04}, \texttt{EPOS-LHC} and \texttt{Sibyll 2.3d} models implemented in CORSIKA.

All events were simulated at $20\degree$ zenith, $180\degree$ azimuth, and with an opening angle of the view cone of $5\degree$.
The choice of these particular zenith and azimuth angles does not lead to a loss in generality, as it is the typical setting for studies evaluating IACTs performances~\cite{HESS:2015cyv,MAGIC:2014zas,CTAConsortium:2017dvg, Glombitza_2023}.
The simulations were performed for the so-called \texttt{phase2d3} configuration, which corresponds to the latest state of the H.E.S.S. array taking into account effects such as the optical degradation of the telescope mirrors and changes in electronics~\cite{ThesisJohannes,ThesisSimon}. Five representative primaries (proton, helium, nitrogen, silicon, and iron) were used. The energy range and total number of events were variable, depending on the primary particle (Table \ref{tab:primaries_en_range}), and a spectral index of -2 was used.
Furthermore, we used the software \texttt{sim\_telarray} (version 1.63) to fully simulate the detector response, ranging from the particles’ ray tracing to the measurement with photomultiplier tubes (PMTs) and its digitization~\cite{Bernl_hr_2008}.

\begin{table}[h!]
    \centering
    \begin{centering}
    \caption{Summary of the energy ranges and the number of triggered showers for each primary particle (which is roughly the same for each hadronic interactions model) together with the final number of events after cuts for the two different configurations considered in this work.}\label{tab:primaries_en_range}
    
            \begin{tabular}{ r r r r r}
                \hline\hline
                 Primary & Energy range [TeV] & \# Triggered showers & \# mono events & \# hybrid events\\ 
                \hline
                Proton & 0.03 - 150 & $\sim 5.1 \cdot 10^7$  & $\sim 3.3 \cdot 10^5$ &$\sim 1.3 \cdot 10^5$  \\
                Helium & 0.03 - 500 & $\sim 3.0 \cdot 10^7$ &$\sim 1.3 \cdot 10^5$ &$\sim 5.0\cdot 10^4$\\
                Nitrogen & 0.04 - 800 & $\sim 3.0 \cdot 10^7$ &$\sim 1.2 \cdot 10^5$ &$\sim 5.0\cdot 10^4$\\
                Silicon & 0.05 - 1000 & $\sim 1.3 \cdot 10^7$ & $\sim 5.3 \cdot 10^4$&$\sim 2.3\cdot 10^4$\\
                Iron & 0.06 - 1200 & $\sim 1.3 \cdot 10^7$ &$\sim 5.2 \cdot 10^4$ &$\sim 2.3\cdot 10^4$ \\

                \hline
            \end{tabular}
            
    \end{centering}
\end{table}

Once the simulations are recorded, a calibration and cleaning procedure is needed. Both were done using the standard H.E.S.S. Analysis Program (HAP); in particular, a dual threshold cleaning procedure is applied to exclude all pixels without a shower signal and to keep only pixels with shower signals in local proximity. 
In this paper, we used the so-called 5/10 tailcuts cleaning~\cite{hess_crab} for CT1-4. For CT5, we use a similar cleaning, but considering two neighboring pixels, 9/16NN2.

\begin{figure}
    \centering
    \includegraphics[width=0.98\textwidth]{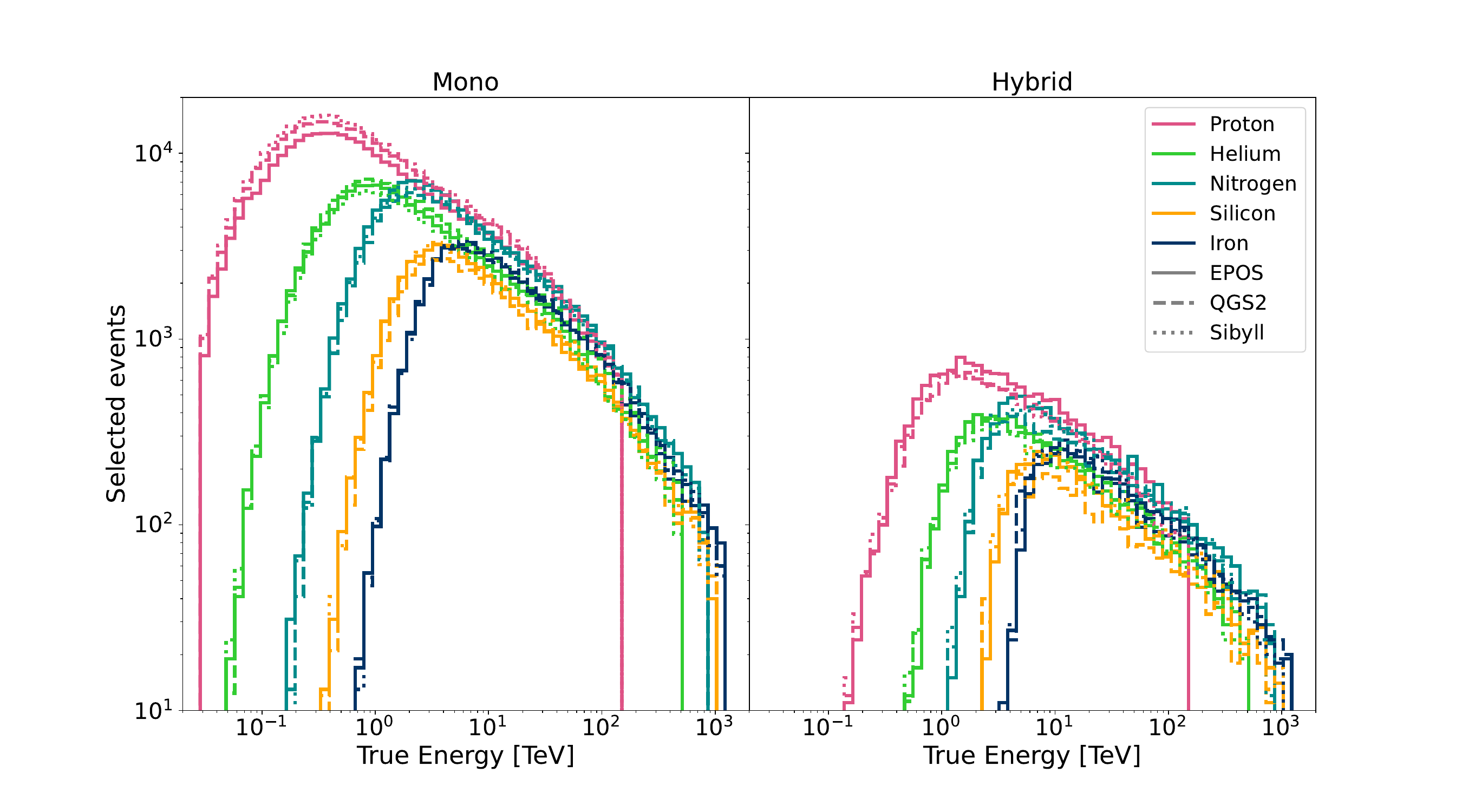}
    \caption{Number of selected events as a function of Monte Carlo true energy for each primary and hadronic model. Selected events are those that triggered the system and survived the preselection cuts. Each color represents a primary, while each line style represents a different hadronic model. Left and right panels are for different preselection cuts: mono, which only considers CT5, and hybrid, which considers CT1-5.}

    \label{fig:simulations}
\end{figure}

We further use pre-selection cuts, part of the standard analysis chain of H.E.S.S., selecting only camera images that are not truncated by the edge of the camera and contain enough light and pixels for a proper Hillas reconstruction. The cuts are listed in Table~\ref{tab:precuts} and ensure high-quality reconstructions even at low energies. The amount of remaining events resulting from the analysis through the different configurations is shown in Table \ref{tab:primaries_en_range} and illustrated in Figure \ref{fig:simulations}.

In H.E.S.S., the reconstruction can be performed with different configurations of the five telescopes enabling studies at different energy ranges. These configurations are motivated by either the start of operations of individual telescopes -- for example -- CT1-4 have been operational since 2003, and CT5 was added to the array in 2012 or by different science goals -- lower energies (from tens to a few hundred of GeV) are only accessible to the CT5 due to its larger size, while higher energy can be detected with the whole array resulting in a higher quality reconstruction. At the time of writing, there are three main configurations: \emph{mono},  \emph{stereo}, and \emph{hybrid}. ``Mono'', corresponds to reconstructions that solely include CT5 -- most effective at low energies. The ``stereo'' and ``hybrid'' configurations correspond to any combination of at least two telescopes from CT1-4 and CT1-5, respectively, having good quality data, i.e., satisfying the required selection. The analyses presented here are done with the mono and hybrid configurations, where the latter has an additional cut applied in order to keep only events detected in all telescopes.

\begin{table}[]
    \centering
    \caption{Preselection cuts used in this analysis. Image amplitude refers to the summed light of the pixels surviving the cleaning and local distance to the distance of the ellipse's center of gravity to the center of the camera. Multiplicity refers to the minimum number of telescopes surviving preselection cuts for the event to be considered.} 
    \label{tab:precuts}
    \begin{tabular}{c|c|c|c|c|c|c|c}
    \hline\hline
        \multirow[c]{2}{*}{Name} & \multicolumn{2}{|c|}{Amplitude [p.e.]} & \multicolumn{2}{|c|}{Local distance [m]} & \multicolumn{2}{|c|}{\# pixels} & \multirow[c]{2}{*}{Multiplicity} \\
        \cline{2-7}
        & CT1-4 & CT5 & CT1-4 & CT5 & CT1-4 & CT5 &  \\
        \hline
        Mono & - & $\geq 80$ & - & $\leq 0.8$ & - & $\geq 5$ & $= 1$ \\
        Hybrid & $\geq 60$ & $\geq 250$ & $\leq 0.525$ & $\leq 0.72$ & $\geq 1$ & $\geq 10$ & $=5$\\
        \hline
    \end{tabular}
\end{table}


\section{Results from low-level variables}\label{sec:lowvariables}
To search for differences in the IACT images resulting from the different hadronic models, first the effects on low-level variables were investigated.
With those, we indicate some parameters directly related to the detected images in the cameras of the IACTs. In particular, in this work, we considered the Hillas parameters \cite{Hillas1985CerenkovLI, crab_hillas_1989ApJ} such as width, length, size (i.e. the total amplitude of the image), kurtosis, and skewness.

Furthermore, we also consider the number of pixels that survive the cleaning process and the so-called ``density'', $D$, and ``length over size'', $L/S$, defined as follows:
\[
D =\frac{\mathrm{Size}}{\mathrm{Width} \cdot \mathrm{Length}}\, , \quad L/S = \frac{\mathrm{Length}}{\log(\mathrm{Size})}.
\]
We also explore the parameter space of true energy, i.e. the simulated energy of the primary particle initiating the shower, and true impact parameter, which here refers to the distance of the shower impact core from the center of the array. As the aim of this study is to investigate the differences between the models themselves, here we only focus on true Monte Carlo parameters.

In the following sections, \ref{subsec:low_mono} and \ref{subsec:low_hybrid}, only results for proton are shown, while the results for helium regarding the low-level variables can be found in the Appendix (\ref{app:low}).

\subsection{Mono} \label{subsec:low_mono}
\begin{figure}[]
    \centering
    \includegraphics[width=0.8\textwidth]{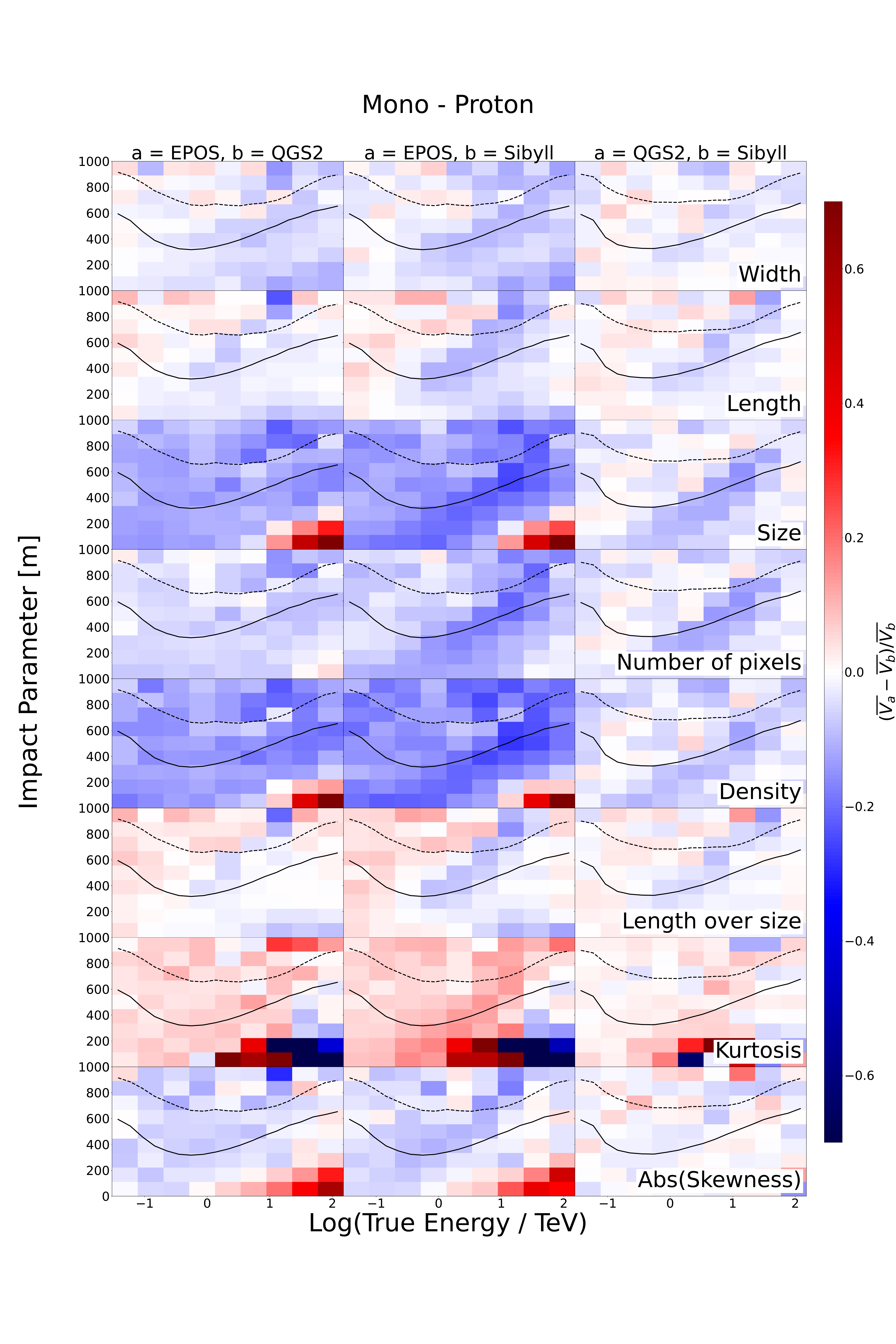}
    \caption{Given a pair of models, a and b, and indicating with V the mean of a given variable, this grid-plot shows the 2D histogram of the $\frac{\overline{V_a} - \overline{V_b}}{\overline{V_b}}$ in impact parameter and true energy bins for the mono configuration. Each row corresponds to a low-level variable, while each column refers to a pair of compared models, specified at the top of the corresponding column. Solid and dashed lines indicate the 68\% and 95\% containment radius for the Monte Carlo true impact parameter. Only proton primaries are shown.}
    \label{fig:2dhist-mono-proton}
\end{figure}

For each variable, the quantity $\frac{\overline{V_a} - \overline{V_b}}{\overline{V_b}}$ was calculated, where $\overline{V}$ stands for the mean of the corresponding variable in a region of the parameter space (e.g., in a bin of true energy and true impact parameter) and $a$ and $b$ are two of the three models to be compared.\\
Figure \ref{fig:2dhist-mono-proton} shows the 2D histogram for each variable (the rows of the grid-plot) and for each combination between the three models (the columns of the grid-plot) when having a proton initiating the air shower. When looking at the differences between the models \texttt{QGSJET-II04} and \texttt{EPOS-LHC} or \texttt{Sibyll 2.3d} and \texttt{EPOS-LHC}, for all the variables in the plot, it is possible to note differences, in particular along the solid and dotted lines, which indicate the 68\% and 95\% containment radius for the Monte Carlo true impact parameter, i.e. where most of the events concentrate.
The comparison between \texttt{QGSJET-II04} and \texttt{Sibyll 2.3d} shows also some differences but not as strong as the other cases and only for some of the variables. It is clear that, for each subplot, the differences are concentrated in specific regions of the phase space. In particular, for the size, density, kurtosis and skewness parameters it is possible to spot an area of high energy and small impact parameter with large differences. Overall, the variables that show the strongest dependencies from the used hadronic interaction model are size, number of pixels, density, kurtosis, and skewness.

\begin{figure}[]
    \centering
    \includegraphics[width=0.98\textwidth]{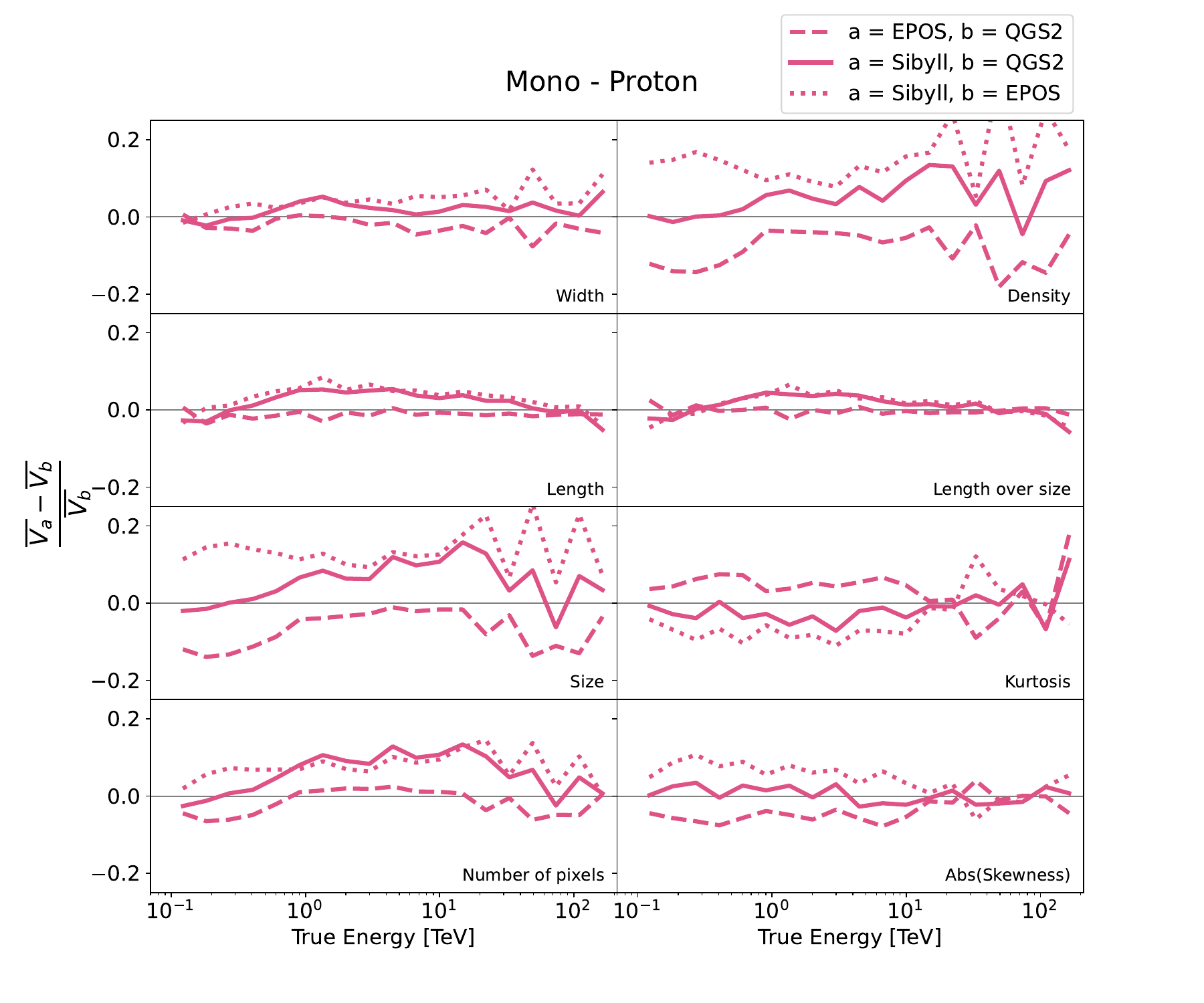}
    \caption{The same quantity for the same set of variables as in Figure \ref{fig:2dhist-mono-proton} is shown here only in energy bins. In each bin, the values are calculated by selecting events around $0.5 \sigma$ of the mean of the impact parameter.  Solid, dashed, and dotted lines indicate the pair of different models being compared, according to the legend on the top right of the plot. Only proton primaries are shown.}
    \label{fig:1d-mono-proton}
\end{figure}

While the different behaviors in different regions of the phase space may allow for a more targeted search, certain regions may suffer from a lack of statistics. In order to inspect the differences that could still be statically significant, we have calculated the quantity $\frac{\overline{V_a} - \overline{V_b}}{\overline{V_b}}$ for all variables, but binned in true energy and by selecting events that are in the range of $\pm 0.5 \sigma$ from the mean of the impact parameter in that energy bin. This is shown in Figure~\ref{fig:1d-mono-proton}.

Here, it is easier to spot which variables are more important in distinguishing the models and which models are more consistent with each other, by looking at how far the lines are from zero. 
The largest effect is found for density, size, and number of pixels. For the first two, the solid line related to the comparison between \texttt{Sibyll 2.3d} and \texttt{QGSJET-II04} does not deviate much for zero. This means that the two models are more in agreement with each other than when these same models are compared to \texttt{EPOS-LHC}.\\

As seen in Appendix~\ref{app:low}, size and number of pixels are also the most significant variables for helium. Nevertheless, the model that differs the most from the other two is \texttt{QGSJET-II04}. The same behavior is found for nitrogen and silicon, while no significant difference is found for iron.
\subsection{Hybrid}\label{subsec:low_hybrid}
\begin{figure}
    \centering
    \includegraphics[width=0.8\linewidth]{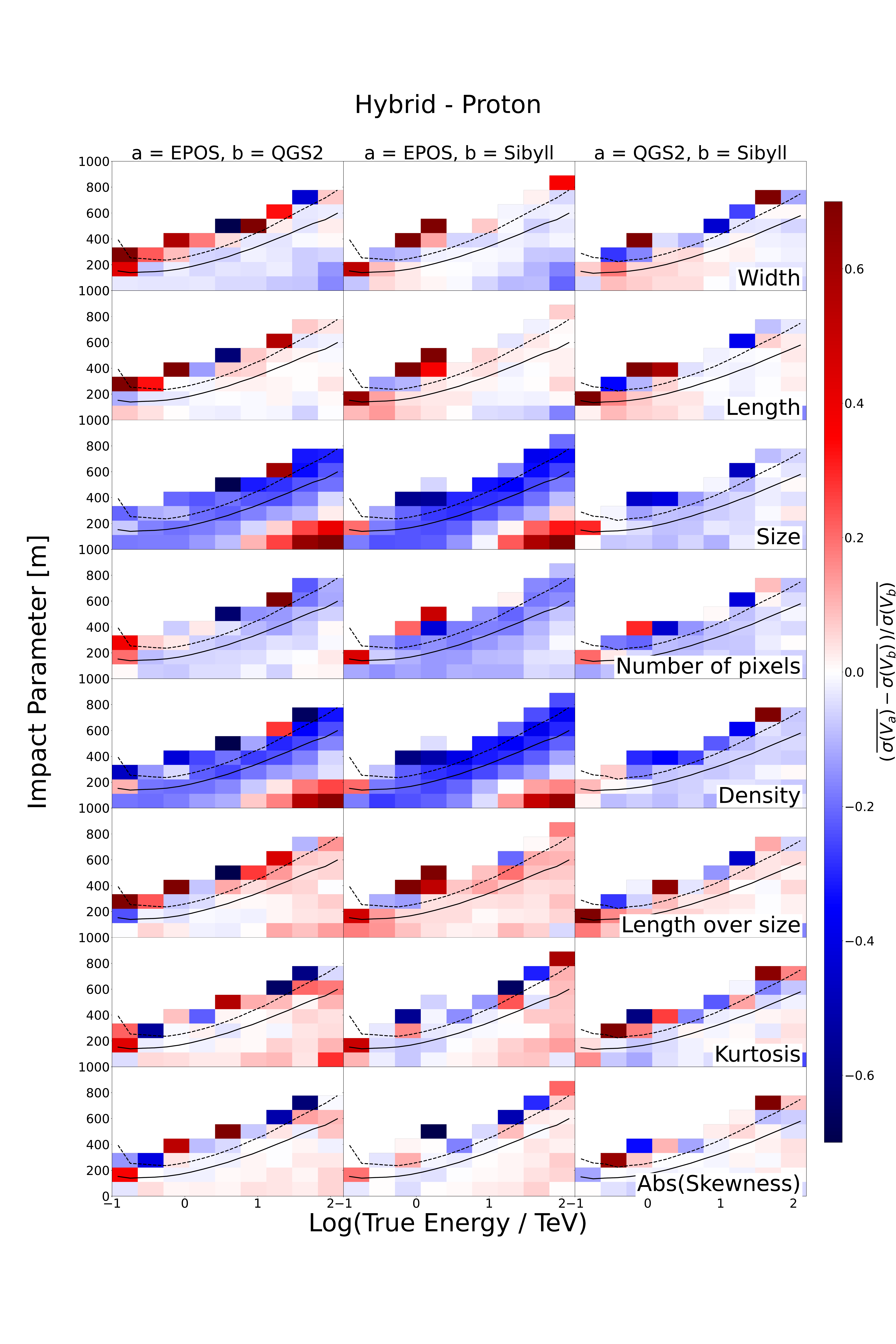}
    \caption{Given a pair of models, a and b, and indicating with V the mean of the standard deviation among CT1-5 of a given variable, this grid-plot shows the 2D histogram of the $\frac{\overline{\sigma(V_a)} - \overline{\sigma(V_b)}}{\overline{\sigma(V_b)}}$ in impact parameter and true energy bins for the hybrid configuration. Each row corresponds to a low-level variable, while each column refers to a pair of compared models, specified at the top of the corresponding column. Solid and dashed lines indicate the 68\% and 95\% containment radius for the Monte Carlo true impact parameter. Only proton primaries are shown.}
    \label{fig:2dhist-hybrid-proton}
\end{figure}

\begin{figure}
    \centering
    \includegraphics[width=0.98\textwidth]{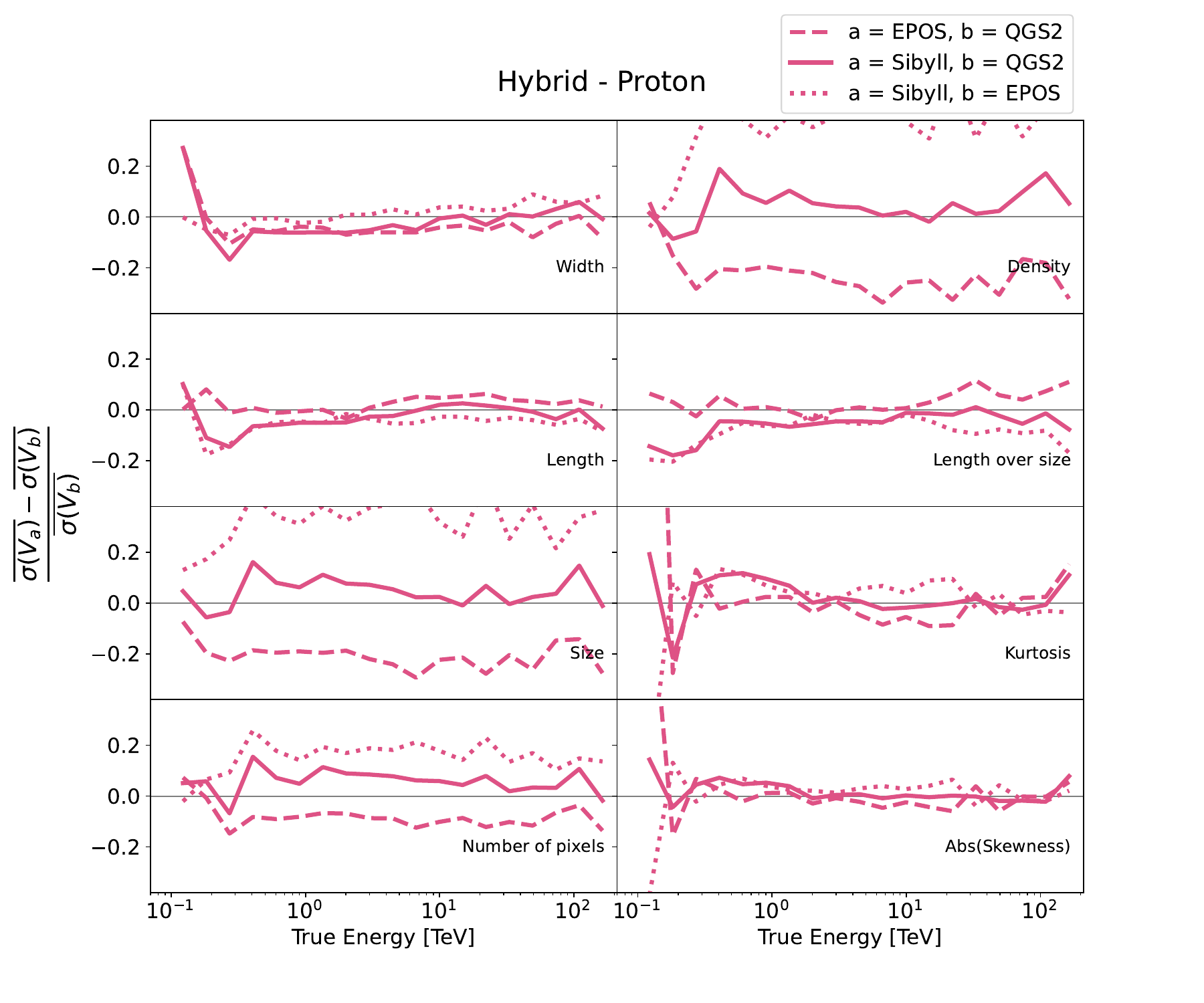}
    \caption{The same ratio for the same set of variables as in Figure \ref{fig:2dhist-hybrid-proton} is shown here only in energy bins. In each bin, the values are calculated by selecting events around $1 \sigma$ of the mean of the impact parameter.  Solid, dashed, and dotted lines indicate the pair of different models being compared, according to the legend on the top right of the plot. Only proton primaries are shown.}
    \label{fig:1d-hybrid-proton}
\end{figure}

Hadronic showers may contain several sub-electromagnetic showers, which result in asymmetric camera images for the different telescopes. This effect, however, is not encapsulated by computing the mean of the variables for each telescope and for each event. Instead, to fully exploit the hybrid configuration, for each variable, the standard deviation among CT1-5 was computed for each event ($\sigma(V)$), then the average among all the events was taken to compare one model to the other. The resulting 2D- and 1D-histograms as before are shown in Figure \ref{fig:2dhist-hybrid-proton} and Figure \ref{fig:1d-hybrid-proton}.\\
The 2D-histograms show fewer structures with the hybrid configuration in comparison to the mono configuration, but it is still possible to spot some differences along the 68\% and 95\% containment radius for some variables like density, number of pixels, and size. Again, those differences are more visible only in the first two columns (\texttt{EPOS-LHC} vs \texttt{QGSJET-II04} and \texttt{EPOS-LHC} vs \texttt{Sibyll 2.3d}), highlighting once more \texttt{EPOS-LHC} as the model that differs the most from the others.\\
This is also confirmed by the hybrid 1D-histogram, where the solid line for the same three mentioned parameters (size, density, and number of pixels) is the closest to zero, meaning that \texttt{QGSJET-II04} and \texttt{Sibyll 2.3d} are harder to distinguish.\\
As for the mono configuration, helium results are shown in Appendix~\ref{app:low}. It is harder to draw clear conclusions from these plots, but a hint is given from the width parameter for which it seems that the \texttt{QGSJET-II04} model is the one differing the most from the others, as already shown for the mono configuration when considering helium, silicon and nitrogen. Iron remains the primary for which it is not possible to assert a particular outcome.\\

\section{Results from high-level variables} \label{sec:highvariables}

Once the effects of each individual variable are better understood, a more sensitive investigation of the (in)compatibility of the models can be explored by combining all the variables. In particular, we use the Boosted Decision Trees (BDT) algorithm, a widely used method in astrophysics~\cite{Coadou:2013lca,Ohm:2009nw,Krause:2017erl,Maier:2017wzr} and proven efficient for classification tasks. We have trained a BDT for each combination of: configuration (mono and hybrid), primary (H, He, N, Si, and Fe), and pair of models (\texttt{EPOS-LHC} versus \texttt{QGSJET-II04}, \texttt{EPOS-LHC} versus \texttt{Sibyll 2.3d}, and \texttt{QGSJET-II04} vs \texttt{Sibyll 2.3d}), resulting in $2 \times 5 \times 3 = 30$ different BDT trainings. The data was split, with $90\%$ of it used for training the model and $10\%$ used for evaluating it. These are called \textit{train} and \textit{test} datasets. The eight variables discussed in section~\ref{sec:lowvariables} were used as an input. Density and size were modified to log density and log size given the vast range of magnitudes covered by these variables. To make it easier to check the stability of the BDT for $a$ versus $b$ and the differences between the evaluated datasets, we convolved the BDT output distribution with a function such that the \textit{train} distribution for model $a$ is flat between 0 and 1. In such a way, a dataset that agrees with model $a$ will present a flat BDT output distribution, while a dataset that agrees with model $b$ will present a non-flat distribution peaked around 1.
After the training, the BDT was also evaluated with the full dataset of the third model not involved in the training. 

\subsection{Mono}
\begin{figure}
    \centering
    \includegraphics[width=0.98\textwidth]{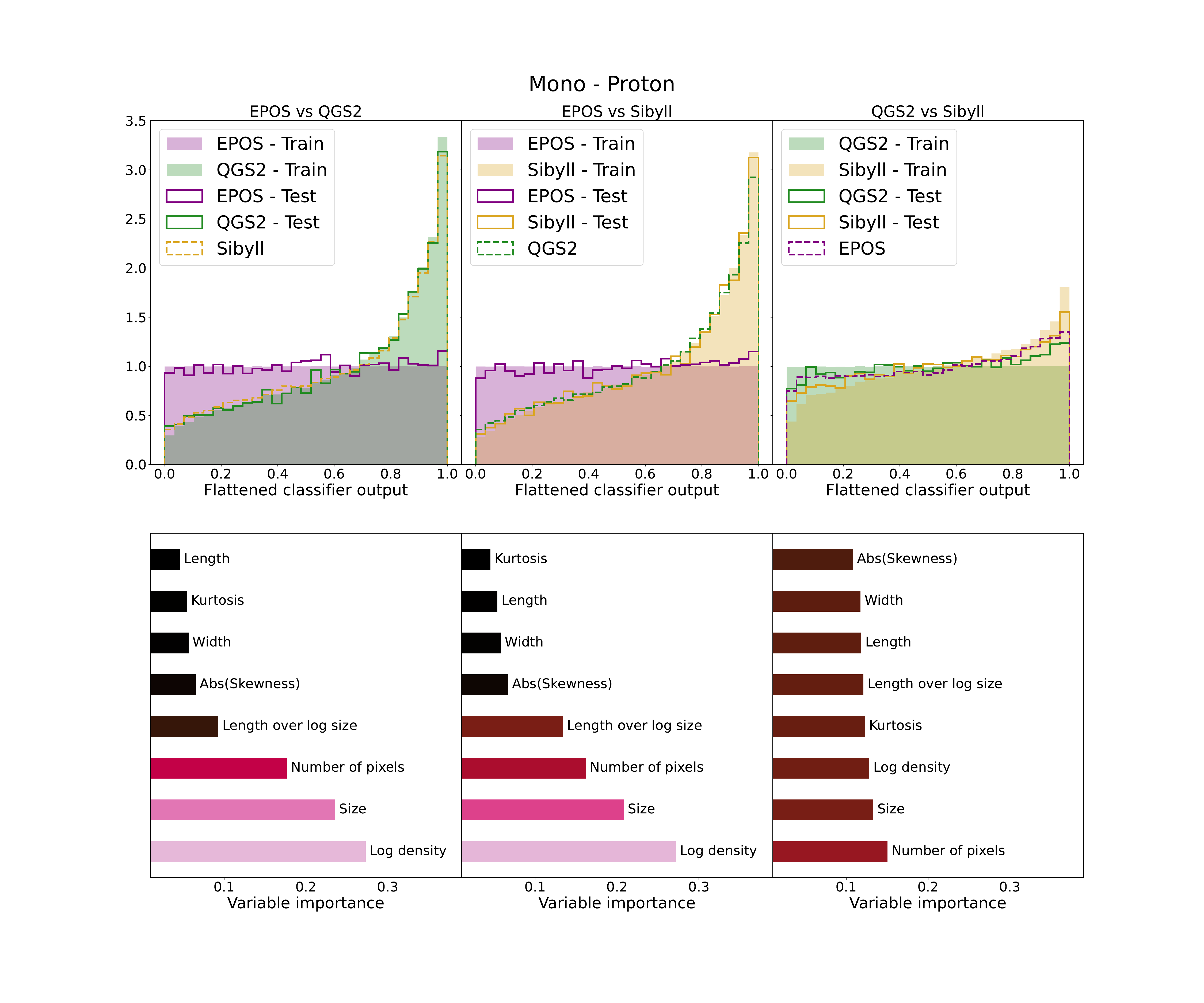}
    \caption{Results of the mono BDT trained for separating each pair of hadronic models. Only primary protons are considered. The top panels show the flattened classifier output distribution for the \textit{train} (shaded areas) and \textit{test} (lines) data as well as for the third model not involved in the training. The bottom panels show the variable importance for each case.}
    \label{fig:BDT-mono-proton}
\end{figure}

Figure~\ref{fig:BDT-mono-proton} shows the results of the BDT for each pair of models for mono and protons. By choice, the distribution for the \textit{train} dataset of one of the models will always be flat. The distribution for the \textit{train} dataset of the other model involved in the training is expected to be significantly different from flat and peak at 1. If that's the case, a successful separation between these models is achieved. Another important indication of the stability of the training is an agreement between the \textit{test} and \textit{train} datasets for each model. The dataset for the third model not involved in the training has also been evaluated in each case. This reveals similarities/differences between the models which are unbiased with respect to any overfitting of the BDT.

With that, it can be noted that the training was stable and successful for \texttt{EPOS-LHC} vs \texttt{QGSJET-II04} and \texttt{EPOS-LHC} vs \texttt{Sibyll 2.3d}. Density, size, and number of pixels are shown to be the most relevant variables, as expected.

For \texttt{QGSJETII-04} versus \texttt{Sibyll 2.3d}, the BDT output distributions are very similar and a bad agreement is found between \textit{test} and \textit{train}. This suggests that the BDT could not find significant differences between these models. A further exploration of the hyperparameters of the BDT, in particular, number of estimators, maximum depth and learning rate, has proven not enough to improve the separation.
This demonstrates even further the similarity between \texttt{QGSJETII-04} versus \texttt{Sibyll 2.3d} models for protons.

\begin{figure}
    \centering
    \includegraphics[width=0.98\textwidth]{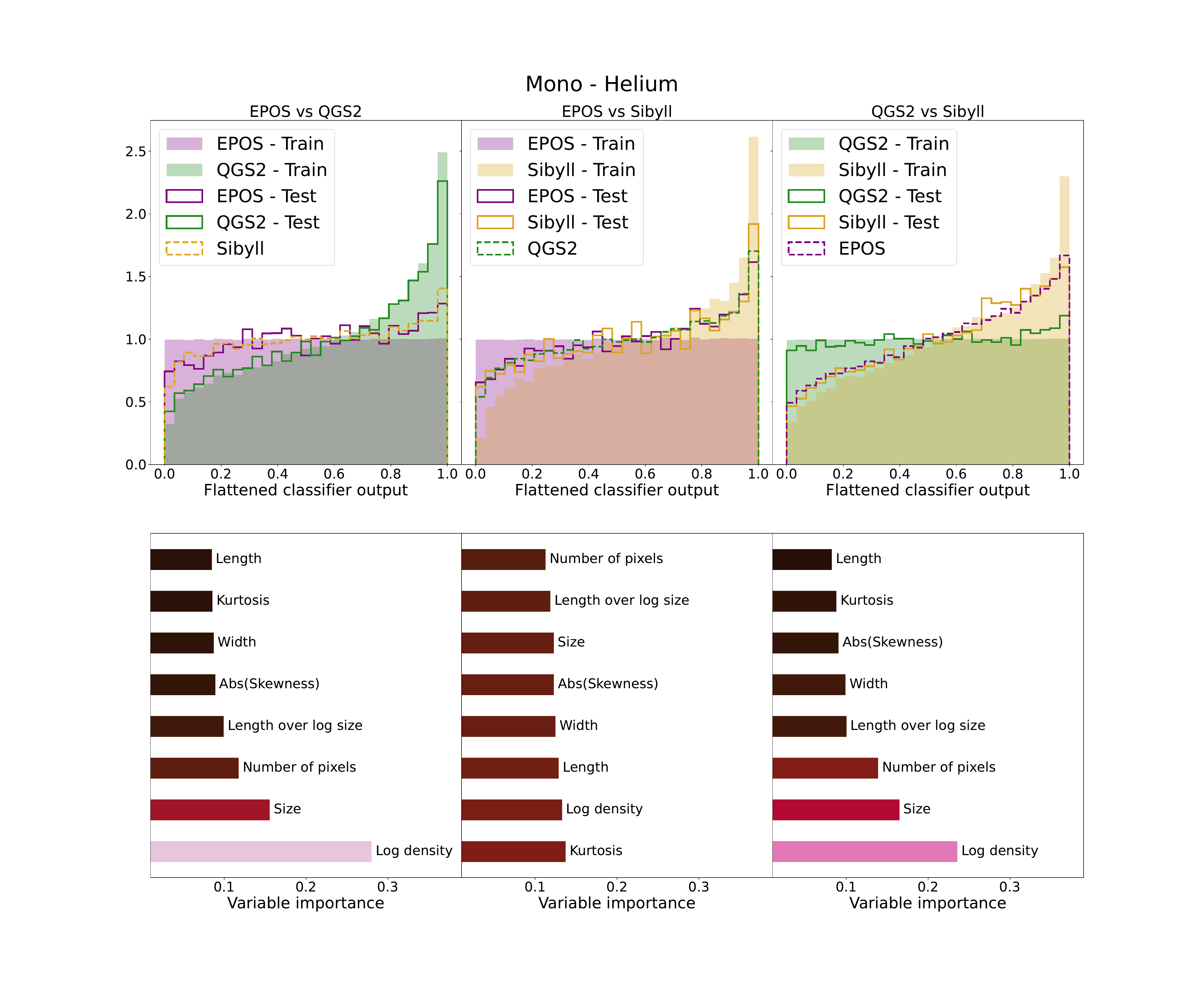}
    \caption{Results of the mono BDT trained for separating each pair of hadronic models. Only primary helium is considered. The top panels show the flattened classifier output distribution for the \textit{train} (shaded areas) and \textit{test} (lines) data as well as for the third model not involved in the training. The bottom panels show the variable importance for each case.}
    \label{fig:BDT-mono-helium}
\end{figure}

\begin{figure}
    \centering
    \includegraphics[width=0.98\textwidth]{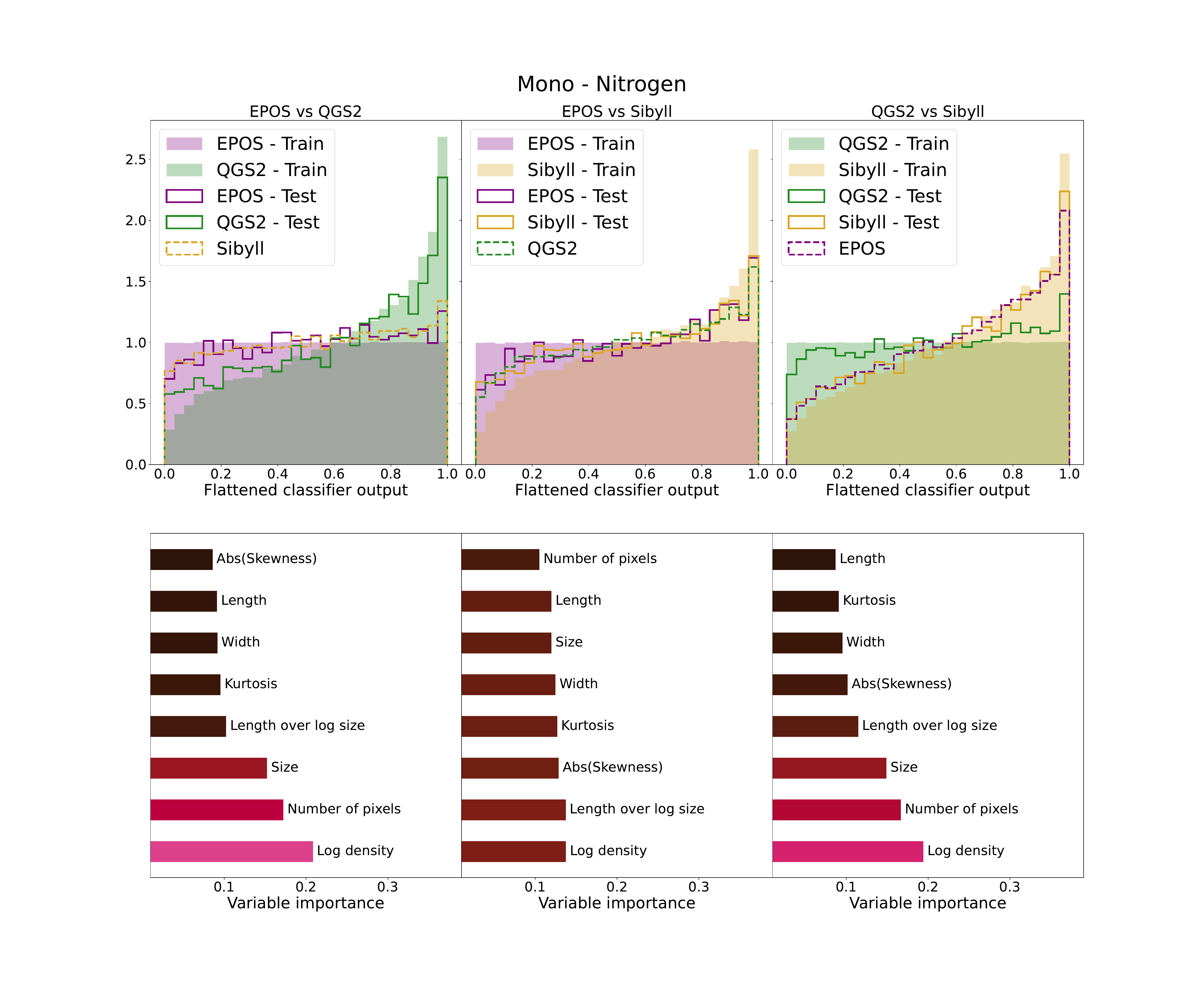}
    \caption{Results of the mono BDT trained for separating each pair of hadronic models. Only primary nitrogen is considered. The top panels show the flattened classifier output distribution for the \textit{train} (shaded areas) and \textit{test} (lines) data as well as for the third model not involved in the training. The bottom panels show the variable importance for each case.}
    \label{fig:BDT-mono-nitrogen}
\end{figure}

Figures~\ref{fig:BDT-mono-helium} and \ref{fig:BDT-mono-nitrogen} show the results for helium and nitrogen. For these, the situation changes and \texttt{QGSJET-II04} is found to be the most different model. For \texttt{EPOS-LHC} versus \texttt{Sibyll 2.3d}, a significant difference between the BDT distribution for the \textit{test} and \texttt{train} datasets can be seen. Both \textit{test} distributions fall somewhere in the middle between the \textit{train} distribution. This is also true for the distribution coming for the third model (\texttt{QGSJET-II04} in this case). This indicates a strong similarity between these models, in agreement with the findings of section~\ref{sec:lowvariables}. No significant separation is found for silicon and iron.

No large differences between the distributions are found such that an event-by-event classification could be possible. This is expected since the effects of the different models are known to not be of leading order in showers for these energies. Nevertheless, if the shapes of the BDT distributions are significantly different and stable (i.e., with the \textit{test} distribution agreeing with the \textit{train} distribution), this effect could be investigated in data by looking at the distribution of events.

A more targeted search could be performed if the differences could be reduced to specific regions of the phase space. We further investigate this by choosing the outlier model for each case (\texttt{EPOS-LHC} for protons and \texttt{QGSJET-II04} for helium and nitrogen) and dividing their \textit{test} data set into two: low BDT, for flattened BDT output < 0.35 and high BDT, for flattened BDT output > 0.65, i.e., the 35\% of events with lowest and highest BDT outputs. Events with low/high BDT output are those in which a high/low separation was found and are, thus, more incompatible/compatible with the other two models. Finally, we again divide the data into different regions of the phase space by binning in true energy and true impact parameter as discussed before. For each bin, we calculate the value of $\xi = (n_{\rm{low BDT}} - n_{\rm{high BDT}}) / (n_{\rm{low BDT}} + n_{\rm{high BDT}})$ as an indicator of incompatibility, with positive/negative values indicating higher/lower incompatibility.

\begin{figure}
    \centering
    \includegraphics[width=0.98\textwidth]{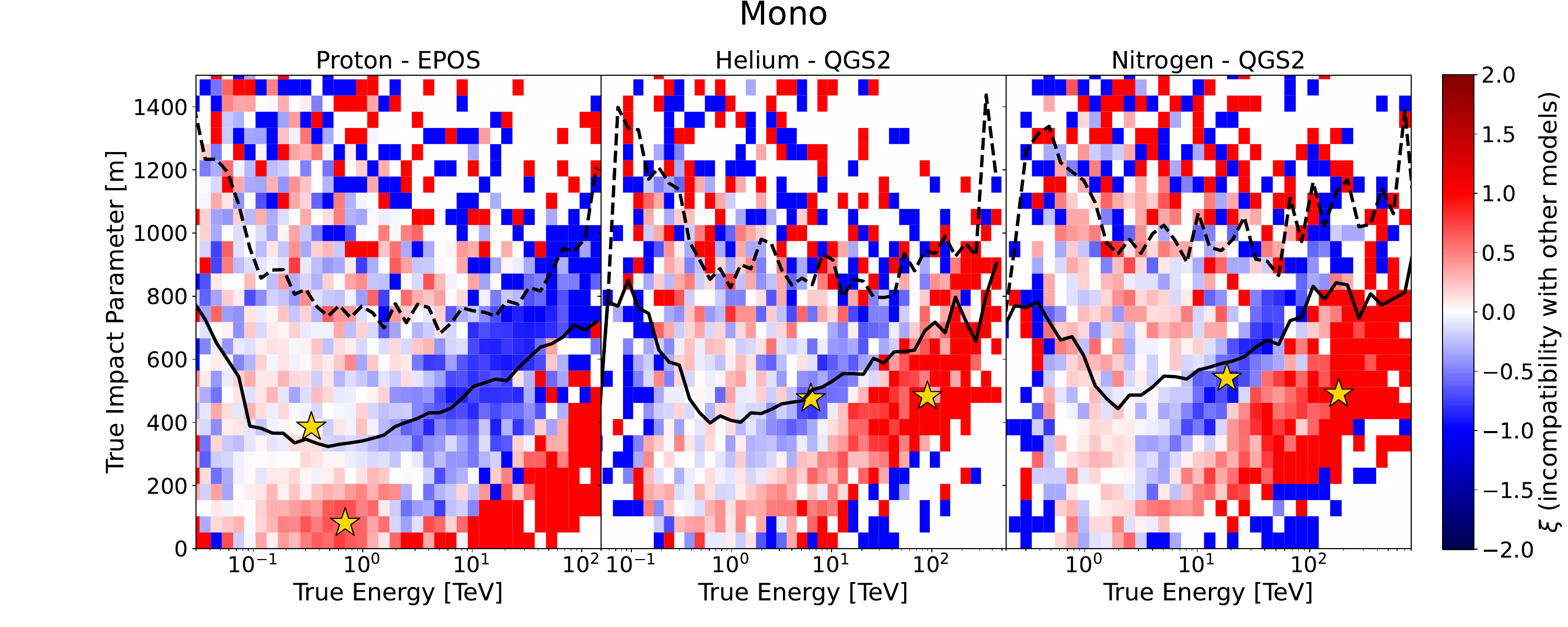}
    \includegraphics[width=0.98\textwidth]{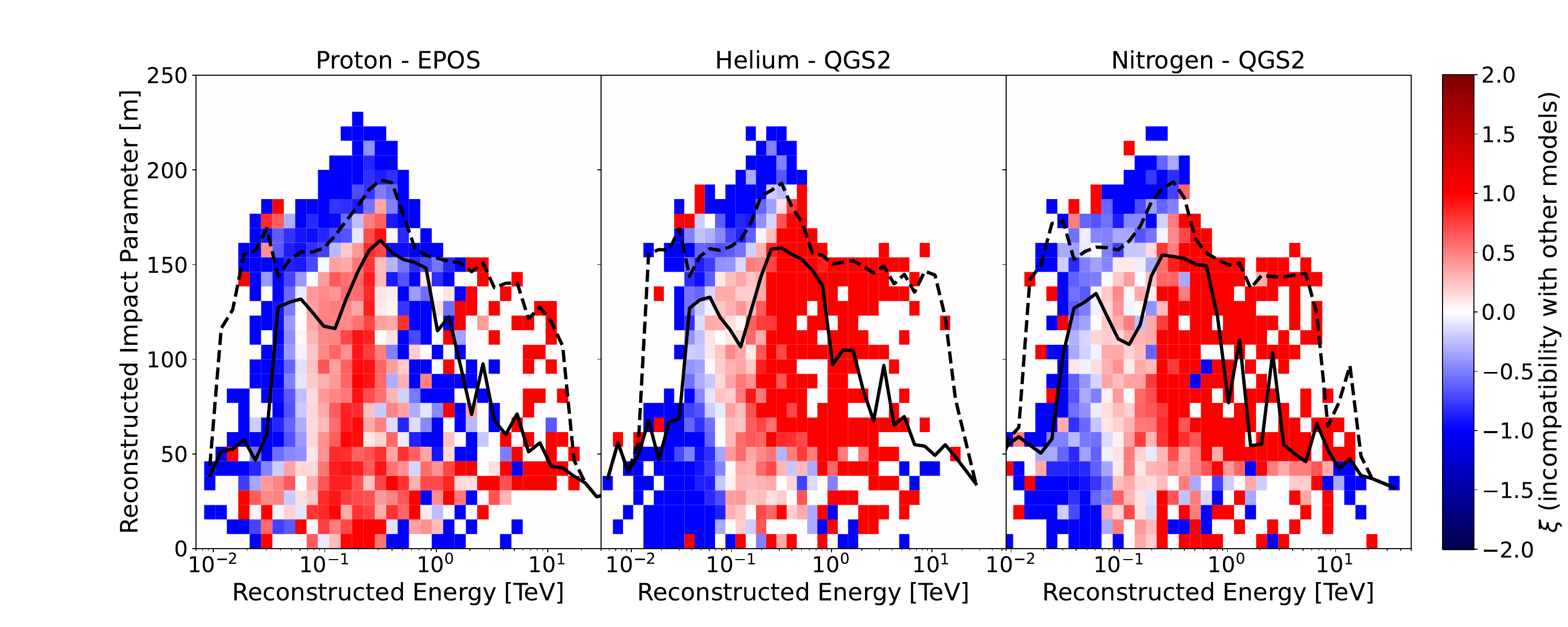}
    \caption{Energy and impact parameter distribution of the (in)compatibility with other 
 hadronic models for a mono configuration. The color scale represents the value of $\xi = (n_{\rm{low BDT}} - n_{\rm{high BDT}}) / (n_{\rm{low BDT}} + n_{\rm{high BDT}})$, for which large positive (red) values mean incompatibility with the other two hadronic models, while large negative (blue) values mean compatibility with the other two hadronic models. The three panels show the results for proton, helium, and nitrogen, respectively. For each primary, the model whose BDT result outliers from the other two is shown. The full and dashed black lines show the 68\% and 95\% containment radius for the Monte Carlo true impact parameter. The yellow stars point out regions of interest further explored in Section~\ref{sec:shower}. The top and bottom panels show the space for true and reconstructed parameters, respectively.}
    \label{fig:Final-mono}
\end{figure}

Here, two different parameter phase spaces are considered. On the one hand, we use true Monte Carlo parameters, which are important for pointing out where are the differences coming from in the models themselves. In this case, we select a few regions of interest (marked in figure~\ref{fig:Final-mono}) to further investigate in the following section.
On the other hand, the phase space of reconstructed parameters is explored. This is crucial as future investigations using data can only be performed with reconstructed parameters. It is, though, important to note that IACT reconstruction is designed for gamma-rays and, for that reason, it is not expected that the distributions of reconstructed and true parameters should agree. This disagreement is more significant for mono, which uses lower energy events and no stereo reconstruction. Nevertheless, this is not a drawback of the analysis, since the goal is to understand which region of the phase space for data is interesting, which will be consistent even if the reconstructed values do not reflect the real ones.

The distributions of $\xi$ are shown in Figure~\ref{fig:Final-mono}. A clear region of incompatibility can be seen in the phase space, with the most incompatible events being those for which the core of the air shower was located closer to the center of the array (correspondingly closer to CT5) in the true parameter case.  These events are mostly within the 68\% containment radius of the impact parameter, showing that they can be explored with the bulk of the statistics without the need to stretch the analysis for very uncommon events. The increase in impact parameter for higher energies probably comes from the fact that, for the highest energies, very low impact parameter events will be truncated and therefore discarded by the preselection cuts.

Six representative regions of the true parameters phase space are chosen to be further investigated in Section~\ref{sec:shower}.

\subsection{Hybrid}

The same procedure was repeated for the hybrid configurations, but now using the standard deviation of the variables as input.

\begin{figure}
    \centering
    \includegraphics[width=0.98\textwidth]{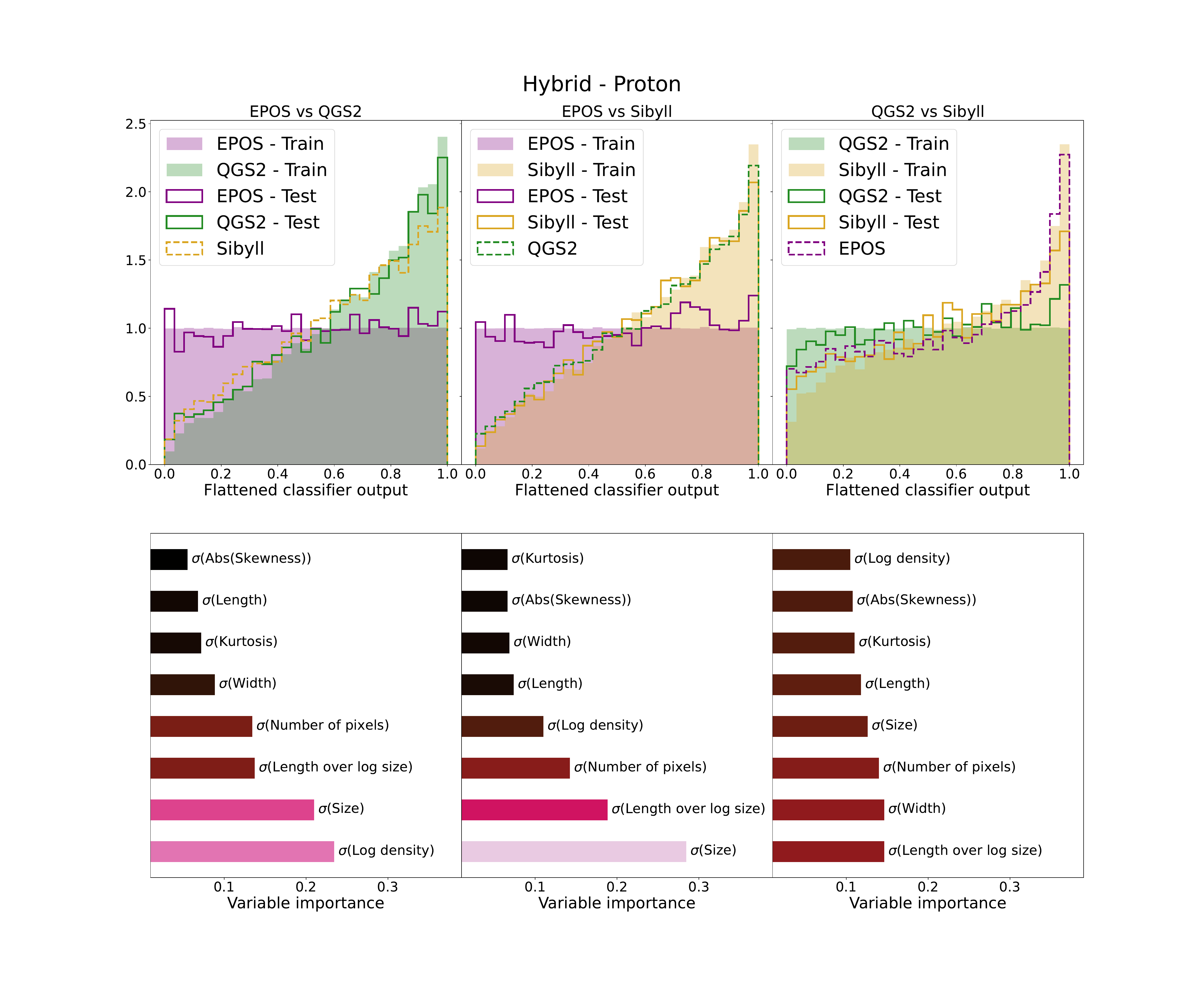}
    \caption{Results of the hybrid BDT trained for separating each pair of hadronic models. Only primary protons are considered. The top panels show the flattened classifier output distribution for the \textit{train} (shaded areas) and \textit{test} (lines) data as well as for the third model not involved in the training. The bottom panels show the variable importance for each case.}
    \label{fig:BDT-stereo-proton}
\end{figure}

\begin{figure}
    \centering
    \includegraphics[width=0.48\textwidth]{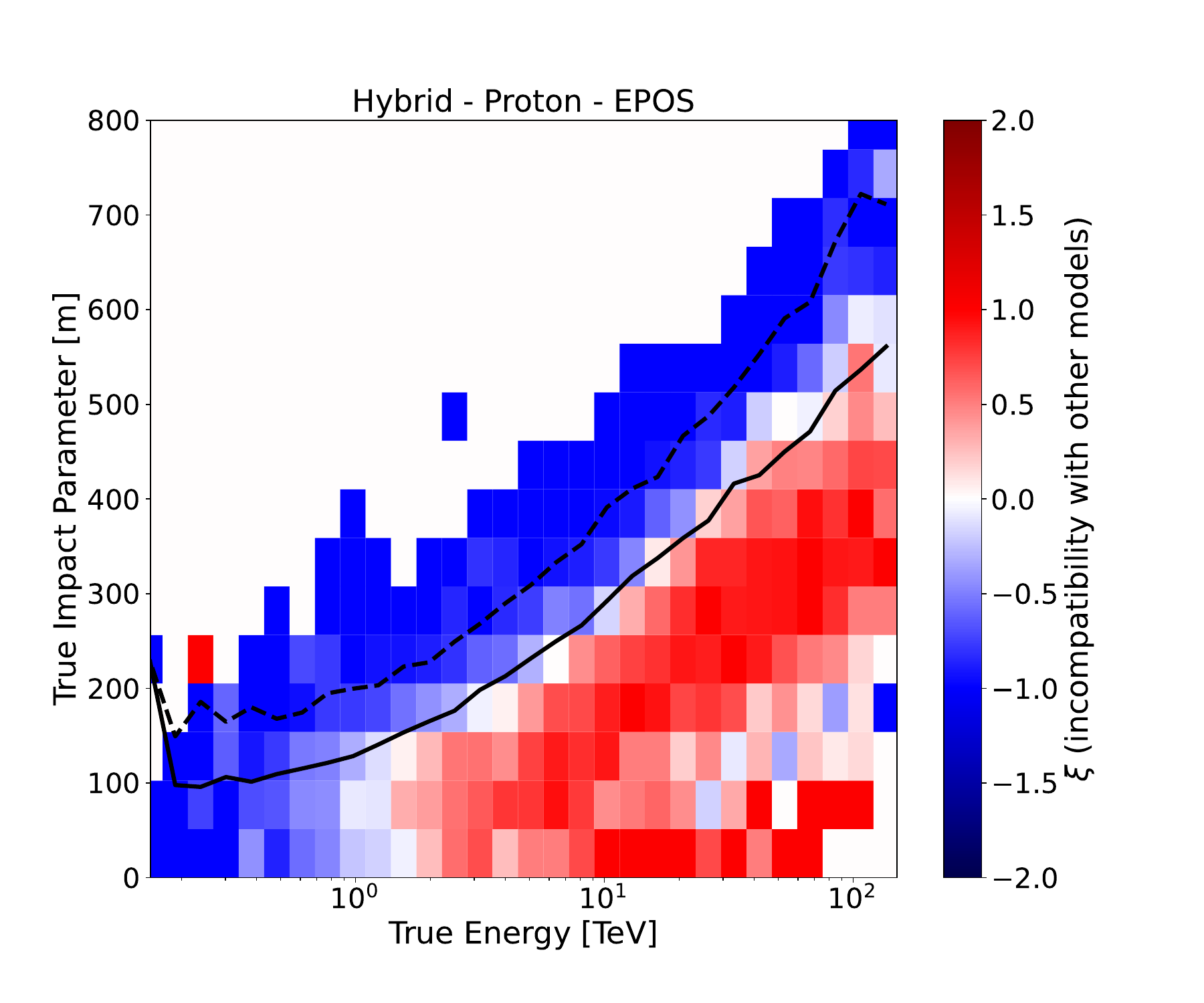}
    \includegraphics[width=0.48\textwidth]{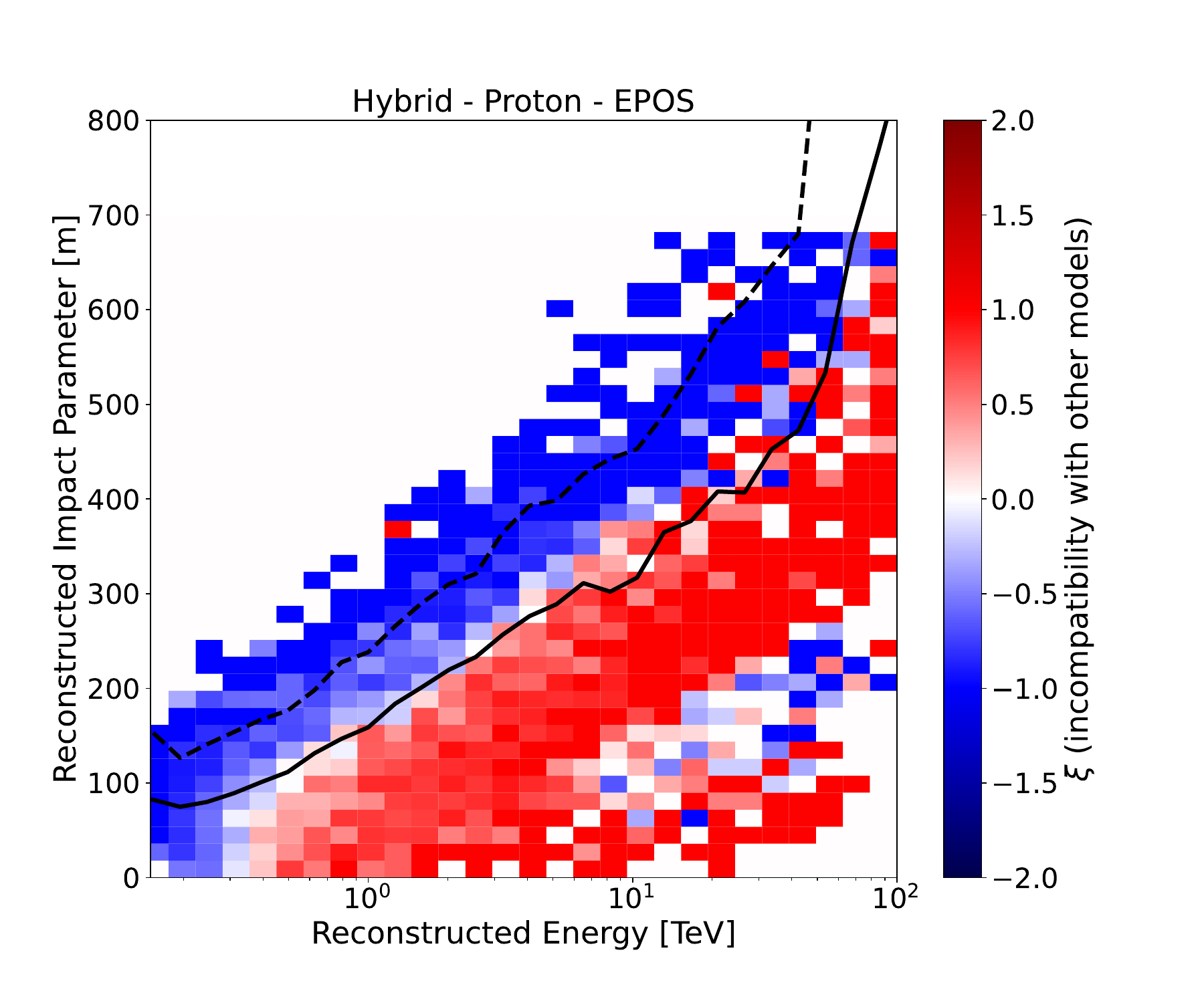}
    \caption{Energy and impact parameter distribution of the (in)compatibility with other 
 hadronic models for a hybrid configuration. The color scale represents the value of $\xi = (n_{\rm{low BDT}} - n_{\rm{high BDT}}) / (n_{\rm{low BDT}} + n_{\rm{high BDT}})$, for which large positive (red) values mean incompatibility with the other two hadronic models, while large negative (blue) values mean compatibility with the other two hadronic models. The three panels show the results for proton, helium, and nitrogen, respectively. For each primary, the model whose BDT result outliers from the other two is shown. The full and dashed black lines show the 68\% and 95\% containment radius for the Monte Carlo true impact parameter. The left and right panels show the space for true and reconstructed parameters, respectively.}
    \label{fig:Final-stereo}
\end{figure}

Figure~\ref{fig:BDT-stereo-proton} shows the results for proton. Once more \texttt{EPOS-LHC} is found to be the outlier model. No significant difference between the models is found for the other primaries. This is expected due to the strict preselection cuts we have chosen for the hybrid configuration.

Finally, figure~\ref{fig:Final-stereo} shows the distribution of incompatibility in the parameter phase space. A clear behavior is seen, with most incompatibility for \texttt{EPOS-LHC} being found in events with small impact parameters. This is consistent with the results found for the mono configuration.

\section{Correlations between shower and Hillas parameter distributions}\label{sec:shower}

With the understanding of the most affected parameters of the IACT images and the corresponding regions of phase space of the shower where larger inconsistencies are found, the next natural step is to build some insight into the difference in shower physics leading to these discrepancies. To do so, we have selected two points of the phase space for each primary in the mono analysis, one in which a large discrepancy is found and one in which a small discrepancy is found. These are represented by a yellow star in Figure~\ref{fig:Final-mono}. For each point, i.e., fixed energy and impact parameter, $10^4$ showers were simulated for each hadronic interaction model. The species and four momenta of the particles generated in the first interaction, as well as the height of the first interaction, were saved for each shower. From that, it was calculated: (a) multiplicity, the number of particles generated, (b) elasticity, the fraction of the energy carried by the most energetic particle, and (c) fraction of energy carried by neutral pions.

\begin{figure}
    \centering
    \includegraphics[width=0.95\textwidth]{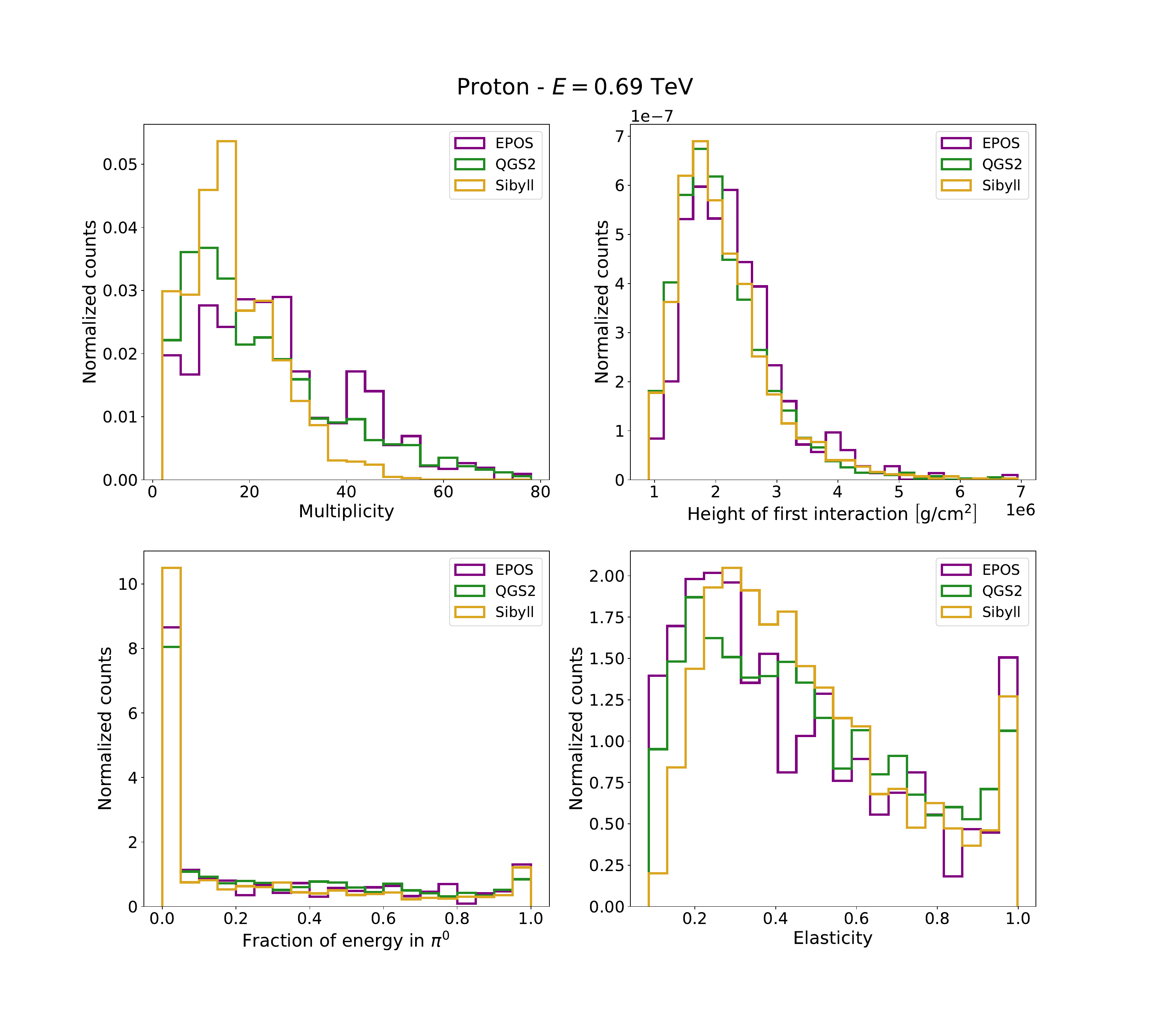}
    \caption{Distribution of parameters of the first shower interaction for the region of interest with large incompatibility for protons. Each color represents a different hadronic model and the panels show distributions for multiplicity, height of first interaction, fraction of energy on $\pi^0$ and elasticity, respectively.}
    \label{fig:showerdistribution}
\end{figure}

Figure~\ref{fig:showerdistribution} shows the distributions for these quantities for the point of interest with larger incompatibility for protons, i.e., $E = 0.69$~TeV. \texttt{EPOS-LHC} was shown in the previous section to be the most discrepant model for this primary. A larger multiplicity and a deeper first interaction are obtained on average with this model. The average elasticity, on the other hand, is smaller, which is expected in an interaction with larger multiplicity, as more particles are present to share the energy.

\begin{figure}
    \centering
    \includegraphics[width=0.95\textwidth]{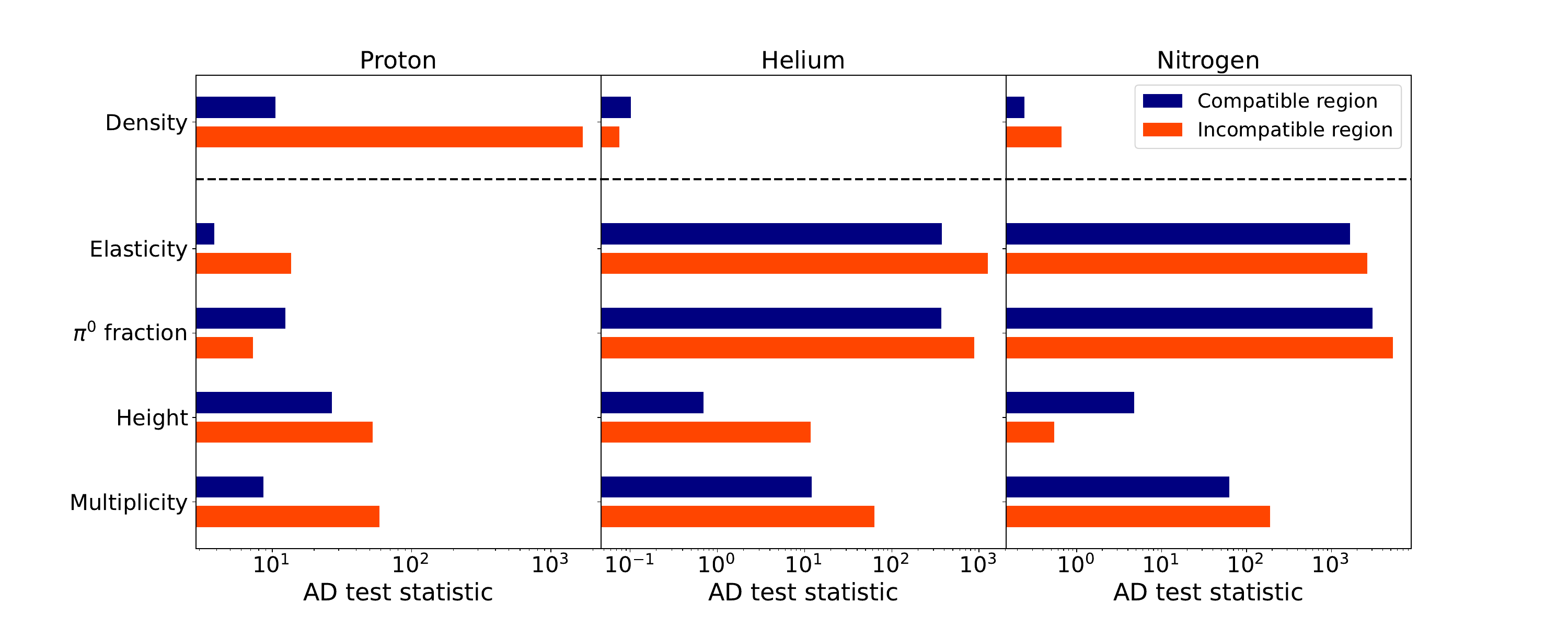}
    \caption{Anderson-Darling statistical test between \texttt{QGSJET-II04}  and \texttt{EPOS-LHC} showers in the same region of the phase space of true parameters. Larger values of the statistical test represent larger intrinsic differences in the distributions. The selected regions are those highlighted by yellow stars in Figure~\ref{fig:Final-mono}, representing a region of high compatibility and a region of high incompatibility for each primary. The dashed line separates Hillas parameters (density) from first shower interaction parameters (elasticity, multiplicity, fraction of energy carried by neutral pions, and elasticity). Each panel represents results for a different primary.}
    \label{fig:adtest}
\end{figure}

A more quantitative estimation of the discrepancies is then investigated by performing an Anderson-Darling test for the distributions of \texttt{EPOS-LHC} and \texttt{QGS-JET-II04} for each of the six points of interest and the four first interaction distributions as well as Hillas density. The results are shown in Figure~\ref{fig:adtest}. For protons, the largest differences are found in the height and multiplicity of the first interaction. For helium and nitrogen, on the other hand, the most significant differences come from elasticity and fraction of the energy carried by neutral pions. As expected, for every significant distribution, a larger discrepancy is seen in the incompatible region with the respect to the one seen in the compatible region.

\section{Conclusions}
\label{sec:conclusions}

Understanding hadronic interactions is a crucial problem for both high-energy physics and astroparticle physics and IACT observatories can be useful in this regard. In this work, we have investigated the possibility of testing hadronic interaction models with IACTs. We performed a range of simulations for extensive air showers with the three most commonly used hadronic models, \texttt{EPOS-LHC}, \texttt{QGSJET-II04} and \texttt{Sibyll 2.3d}, and IACT response using H.E.S.S. as our show-case scenario.

The main focus of the work was on evaluating the effects of the choice of the hadronic model in the IACT camera images, in particular, in the distribution of the Hillas parameters. The work was divided into two main approaches. First, the distributions of different Hillas parameters were investigated independently. Two different configurations, five different primaries and the three possible combinations of each pair of models were evaluated. Number of pixels, size and density were shown to be the variables with the most significant differences. \texttt{EPOS-LHC} showed the largest differences to the other two models for protons and \texttt{QGSJET-II04} for helium, nitrogen, and silicon. No clear difference was seen for iron. These results can shine a light on understanding possible differences coming from the different assumptions made by each of the models considered.

A further investigation was performed by combining all the Hillas parameters into a BDT. Even though an event-by-event classification is not possible (and was not envisioned), a significant and stable difference was found in the BDT distributions for some cases. It was once more verified that \texttt{EPOS-LHC} stands out for protons, while \texttt{QGSJET-II04} stands out for helium and nitrogen. The incompatibility of the models was explored for different regions of the phase space of reconstructed and true parameters. From the reconstructed parameters, it was possible to determine target regions to be explored by further studies using measured data. 

The true phase space was then used to determine interesting regions in which the models were further investigated. Distributions of first interaction parameters were investigated showing that the main 
discrepancies between \texttt{EPOS-LHC} and \texttt{QGSJET-II04} are in the multiplicity and height of first interaction for protons and elasticity and fraction of energy carried by neutral pions for helium and nitrogen.

Therefore, we propose this method as a novel tool for using IACT data to explore the compatibility of the hadronic models with experimental data, allowing further constraining of the models. We also show that a more targeted search could be fruitful by separately analyzing the data in certain regions of the phase space of energy and impact parameter. While H.E.S.S. is used as a show-case scenario, the methods here presented can easily be applied to other IACT experiments as well as to the forthcoming CTAO.

\acknowledgments
We thank the H.E.S.S. Collaboration for allowing us to use H.E.S.S. simulations for this publication. The idea for this work was spawned during a series of two joint FAPESP/BAYLAT workshops titled ``Astroparticle physics in the era of CTA and SWGO'' at the Friedrich-Alexander-Universit\"at Erlangen-N\"urnberg and the Instituto de F\'{i}sica de S\~ao Carlos, Universidade de S\~ao Paulo in 2023 and 2024. We acknowledge the generous support of these workshops by FAPESP (through grant number 2022/01271-7) and BAYLAT. VdS also thanks FAPESP grant number 2021/01089-1. The authors would like to thank Sam Spencer and Alison Mitchell for valuable feedback on the draft of this manuscript. The authors also thank the referee for valuable comments that lead to a deeper investigation of the subject in this work.

\printbibliography
\newpage
\appendix

\section{Appendix} \label{app:low}

\begin{figure}[h!]
    \centering
    \includegraphics[width=0.8\textwidth]{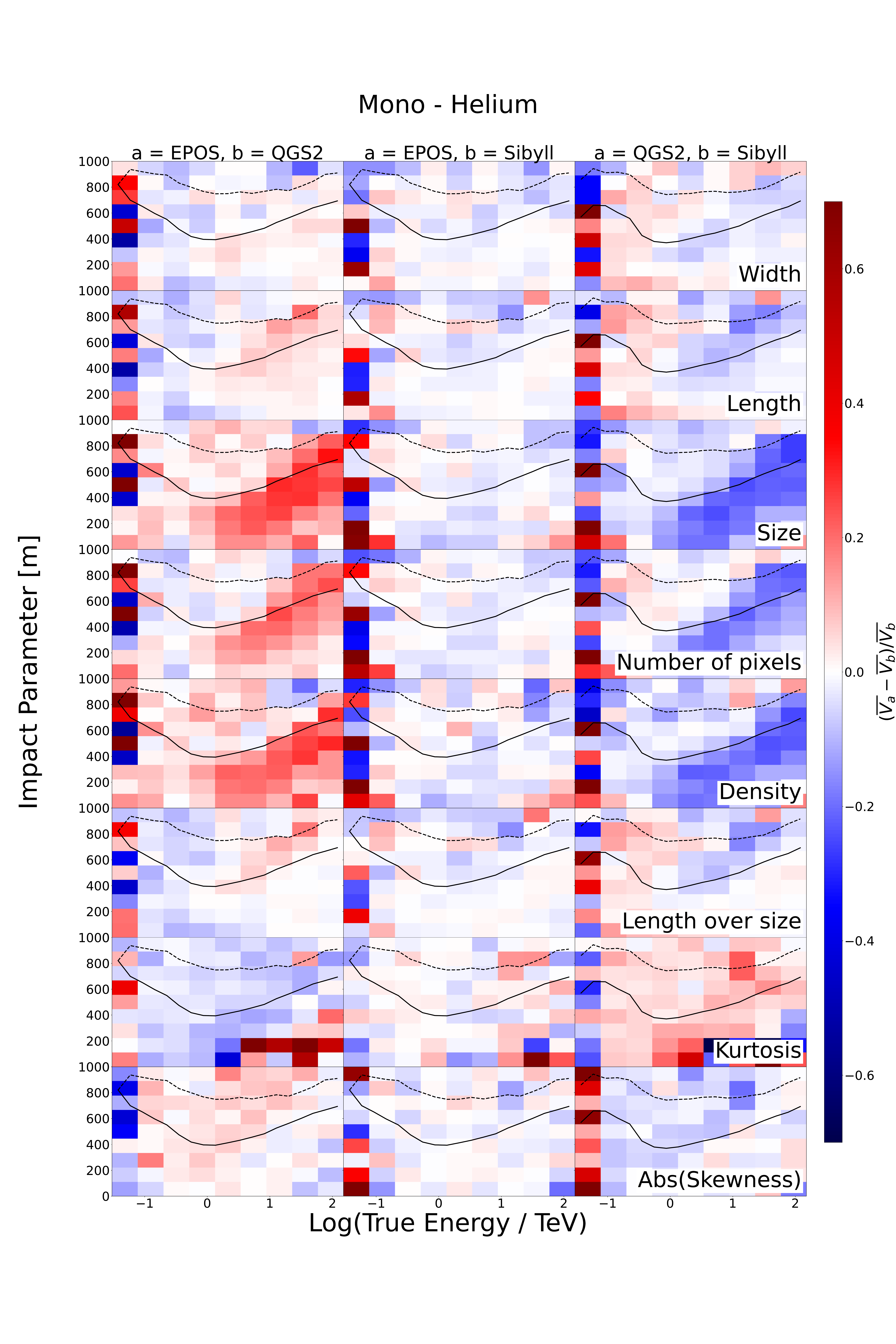}
    \caption{Given a pair of models, a and b, and indicating with V the mean of a given variable, this grid-plot shows the 2D histogram of the $\frac{\overline{V_a} - \overline{V_b}}{\overline{V_b}}$ in impact parameter and true energy bins for the mono configuration. Each row corresponds to a low-level variable, while each column refers to a pair of compared models, specified at the top of the corresponding column. Solid and dashed lines indicate the 68\% and 95\% containment radius for the Monte Carlo true impact parameter. Only helium primaries are shown.}
    \label{fig:2dhist-mono-helium}
\end{figure}
\begin{figure}
    \centering
    \includegraphics[width=0.8\textwidth]{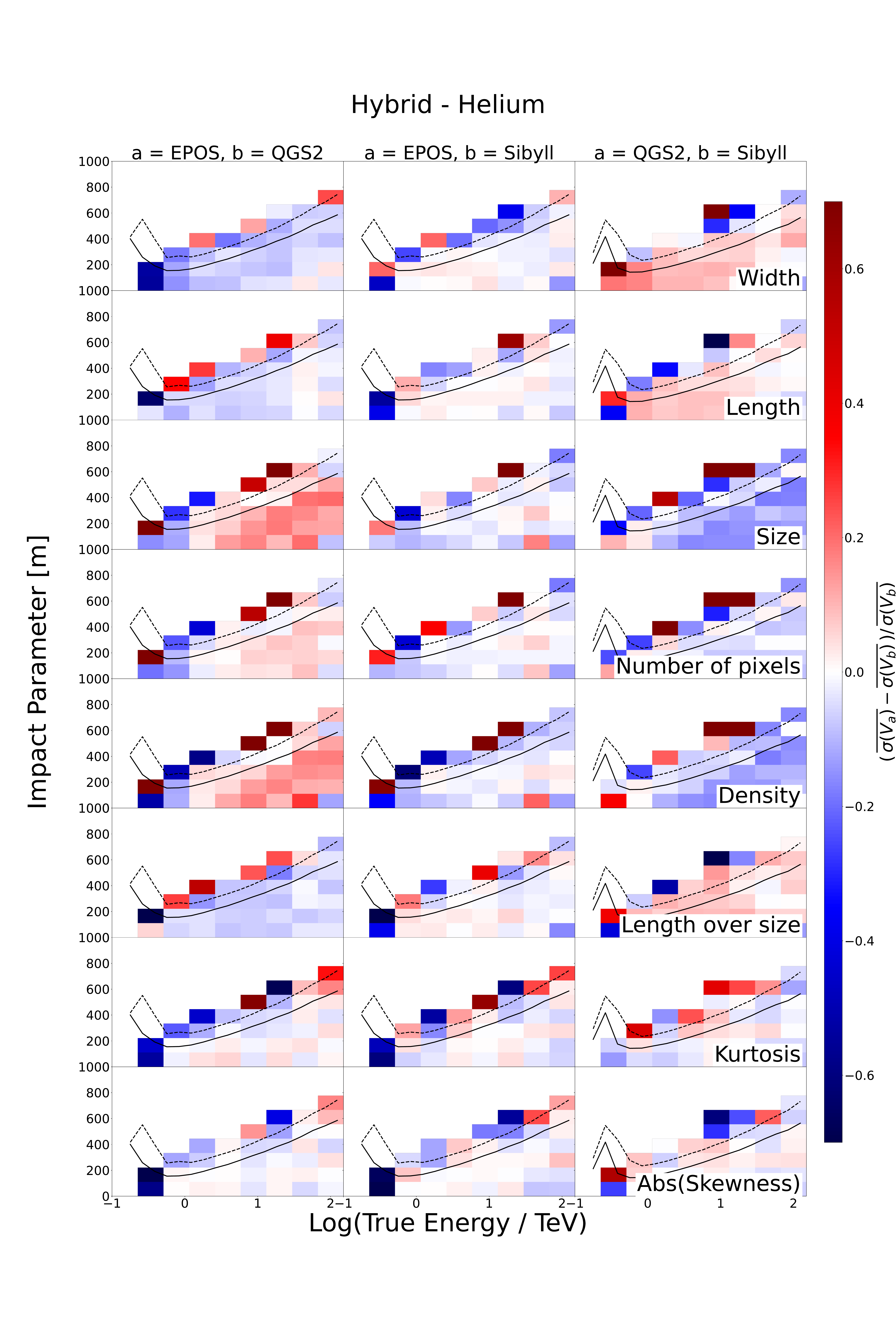}
    \caption{Given a pair of models, a and b, and indicating with V the mean of the standard deviation among CT1-4 of a given variable, this grid-plot shows the 2D histogram of the $\frac{\overline{\sigma(V_a)} - \overline{\sigma(V_b)}}{\overline{\sigma(V_b)}}$ in impact parameter and true energy bins for the hybrid configuration. Each row corresponds to a low-level variable, while each column refers to a pair of compared models, specified at the top of the corresponding column. Solid and dashed lines indicate the 68\% and 95\% containment radius for the Monte Carlo true impact parameter. Only helium primaries are shown.}
    \label{fig:2dhist-hybrid-helium}
\end{figure}

\begin{figure}
    \centering
    \includegraphics[width=0.75\textwidth]{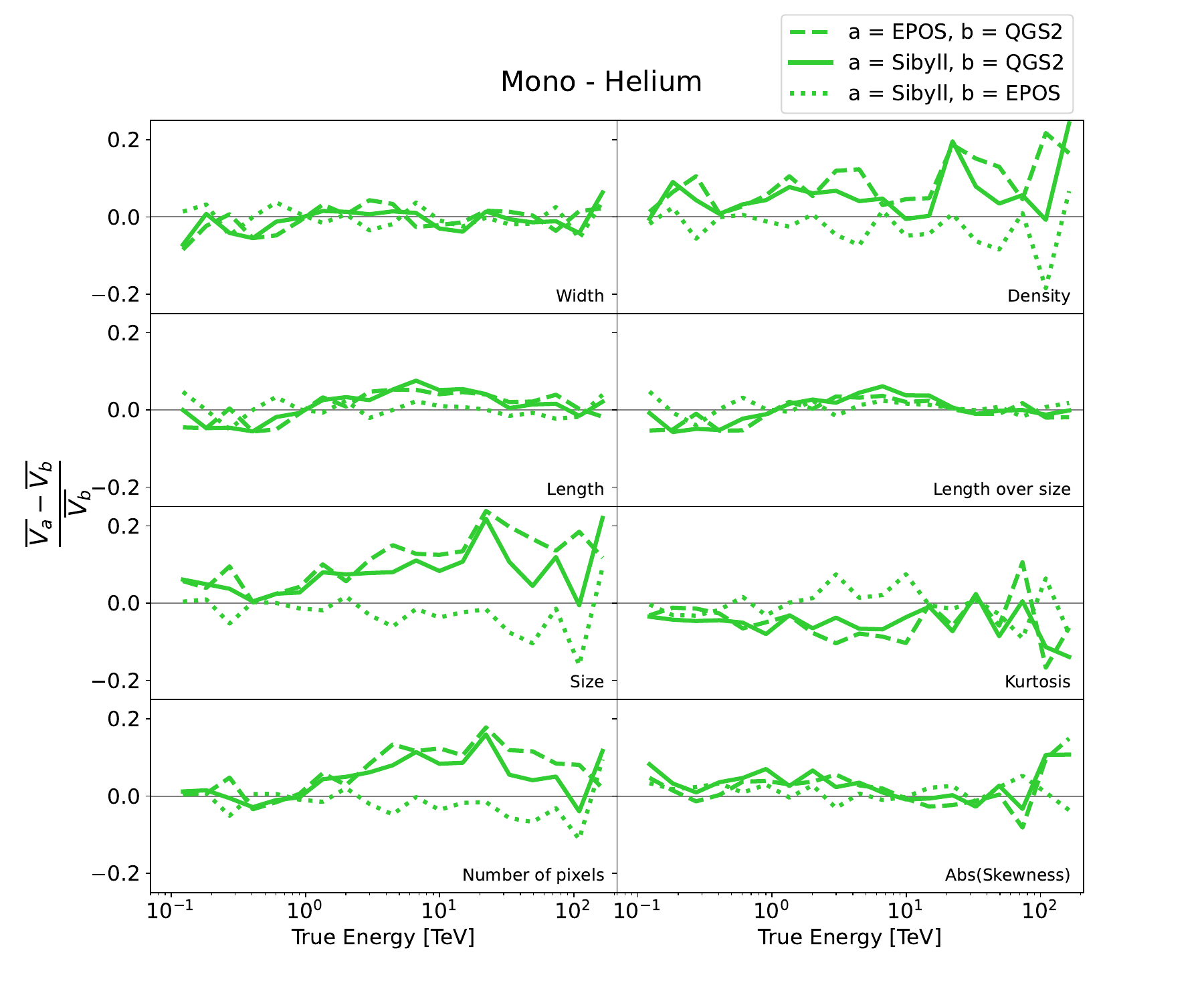}
    \caption{The same quantity for the same set of variables as in Figure \ref{fig:2dhist-mono-helium} is shown here only in energy bins. In each bin, the values are calculated by selecting events around $0.5 \sigma$ of the mean of the impact parameter.  Solid, dashed, and dotted lines indicate the pair of different models being compared, according to the legend on the top right of the plot. Only helium primaries are shown.}
    \label{fig:1dhist-mono-helium}
\end{figure}
\begin{figure}
    \centering
    \includegraphics[width=0.75\textwidth]{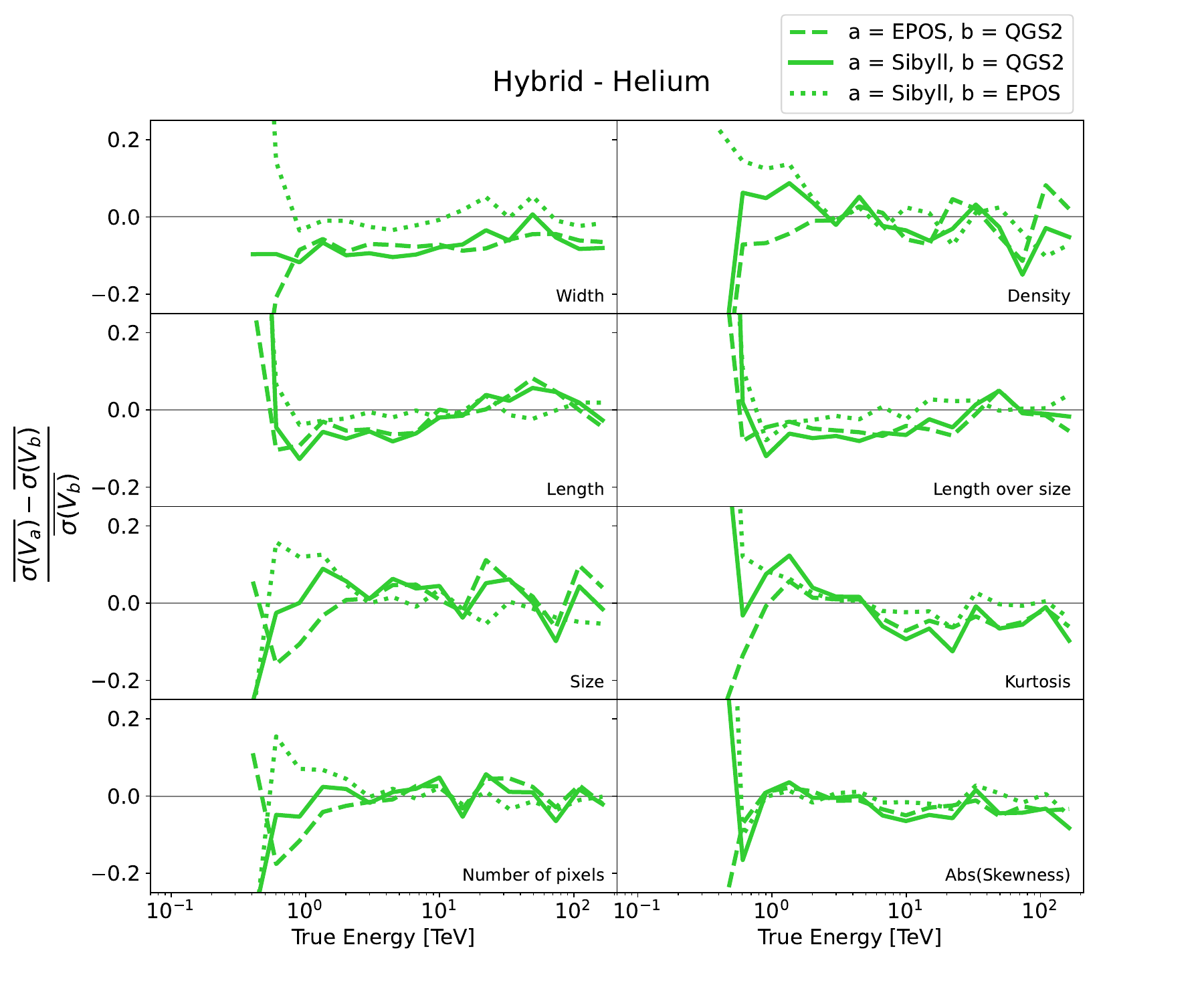}
    \caption{The same ratio for the same set of variables as in Figure~\ref{fig:2dhist-hybrid-helium} is shown here only in energy bins. In each bin, the values are calculated by selecting events around $1 \sigma$ of the mean of the impact parameter.  Solid, dashed, and dotted lines indicate the pair of different models being compared, according to the legend on the top right of the plot. Only helium primaries are shown.}
    \label{fig:1dhist-hybrid-helium}
\end{figure}

\end{document}